\documentstyle[prb,aps]{revtex}        
\tighten

\begin{document}

\title{Domain scaling and marginality breaking in the random-field Ising model}
\author{
E. D. Moore$^a$\footnote{Electronic address: dmoore@thphys.ox.ac.uk},
R. B. Stinchcombe$^a$,
and S. L. A. de Queiroz$^b$\footnote{Electronic address:
sldq@portela.if.uff.br}
 }
\address{
$^a$ Department of Physics, Theoretical Physics, University of Oxford,\\
 1 Keble Road, Oxford OX1 3NP, United Kingdom\\
$^b$Instituto de F\'\i sica, Universidade Federal Fluminense,\\ Outeiro de
S\~ao Jo\~ao Batista s/n, 24210-130 Niter\'oi RJ, Brazil}
\date{\today}
\maketitle
\begin{abstract}
A scaling description is obtained for the $d$--dimensional random field Ising
model from domains in a bar geometry.  Wall roughening removes the marginality
of the $d=2$ case, giving the $T=0$ correlation length $\xi \sim \exp\left(A
h^{-\gamma}\right)$ in $d=2$, and for $d=2+\epsilon$ power law behaviour with
$\nu = 2/\epsilon \gamma$, $h^\star \sim \epsilon^{1/\gamma}$.  Here, $\gamma =
2,4/3$ (lattice, continuum) is one of four rough wall exponents provided by the
theory.  The analysis is substantiated by three different numerical techniques
(transfer matrix, Monte Carlo, ground state algorithm).  These provide for
strips up to width $L=11$ basic ingredients of the theory, namely free energy,
domain size, and roughening data and exponents.
\end{abstract}

\twocolumn

\section{Introduction}

The phase diagram and critical properties of the $d$--dimensional random field
Ising model has long been a subject of great interest
\cite{imr75,gri76,aha76,you77,aha78,aha78b,vil82,gri82,imb84,bri87,aiz89,nat88,you93,gof93}.
 This is partly because of its role~\cite{fis79} in describing the striking
behaviour
\cite{yos82,bel83,sha84,bir85,yos85,mit86,ges88,ram88,fer91,led92,bec93,hil93,hil91}
of diluted antiferromagnets in a uniform field.  However, the model has become
celebrated in its own right because of the way it captures in an extreme way
the effects of randomness and frustration; and because of continuing questions
concerning its lower critical
dimension~\cite{imr75,vil82,gri82,imb84,bri87},
the applicability of dimensional reduction~\cite{aha76,you77,gof93} and the
 nature of the transition and critical behaviour~\cite{aiz89,you93,gof93}.
As is well known, the lower critical dimension $d_l$ turns out to
be~\cite{imb84,bri87} that originally suggested by domain wall
arguments~\cite{imr75,vil82,gri82}, namely $d_l=2$.  This implies a marginal
 behaviour at dimension $d=2$, and it has been
argued~\cite{gri83b,bin83,vil85,nat85} that here domain roughening effects are
important.  This present paper provides a detailed study based on this point of
view.  We explore the consequences of a roughened domain wall picture, and at
the same time confirm its validity by observing characteristics of the basic
ingredients as well as consequent predicted behaviour in data from numerical
studies.

The domain analysis is most conveniently made in a bar geometry.  The
dependence on a bar width $L$ may be used to arrive at phenomenological scaling
tranformations.  These in turn give the phase diagram and critical properties.
In the particular case of low random field strength $h$ and temperature $T$,
the domains span the bar width, and are well separated along the bar, so the
basic ``flat wall'' description (without roughening) is very simple.  As
expected, it results, for $d=2$, in a zero temperature fixed point at which the
field scaling is marginal.  Adding domain wall roughening breaks this
marginality, and this ingredient is essential at and near $d=2$.  Consequences
are: $(i)$ the bulk correlation length behaviour $\xi \sim
\exp\left(A/h^\gamma\right)$ at $T=0$ in $d=2$, and $(ii)$ for $d=2+\epsilon$
the phase boundary joining the $T \neq 0$, $h = 0$ fixed point to one at $T=0$,
$h = h^\star \sim \epsilon^{\left(1/\gamma\right)}$ where $\xi \sim
|h-h^\star|^{-2/\epsilon\gamma}$.  $\gamma$ is one of four exponents occurring
in the $h$ and $L$ dependence of the wall roughening free energy and
characteristic scale, and is predicted to be $\gamma = 2$ in the two
dimensional lattice.  These results have been briefly reported
elsewhere~\cite{intro}.

This description has been here tested by direct numerical investigations in the
bar geometry for $d=2$ (strips).  This is carried out by transfer matrix
techniques (another reason for using the bar geometry) and by Monte Carlo
analysis using a new thermalization technique~\cite{mymc}.  In addition, we use
data from a max--flow algorithm for constructing ground states~\cite{ogi86a}
which has been adapted for strips.

The domain description involves free energy ${\cal F}$ and domain size $\Xi$ as
well as roughening characteristics.  The largest eigenvalue of the transfer
matrix gives accurate numerical data for ${\cal F}$, and both ${\cal F}$ and
$\Xi$ are available from the Monte Carlo analysis, for comparison with the
theory.  The wall roughening affects both ${\cal F}$ and $\Xi$, but its
characteristics are most directly seen in measures of the wall profile itself.
These and their associated exponents are best provided by the ground state
algorithm.  The resulting comparisons of theory and numerical data give a very
complete test of the theoretical description and convincing support for its
validity.

The outline of the paper is as follows.  In section~\ref{sec:fwtheory} the
domain picture is introduced and the flat wall theory is provided at $T=0$ for
the free energy, domain size, and correlation length of bars, and via
phenomenological scaling, for the bulk criticality. Section~\ref{sec:lot}
generalises the flat wall description to low temperatures, and
section~\ref{sec:fwtest} describes the numerical approaches and the comparison
of free energy and domain size data with flat wall theory.
Section~\ref{sec:rough-theory} describes the domain wall roughening at $T=0$,
for lattice and continuum models via a simple approach and a field theory.
First, a single ``decoration'' is discussed, then decorations on all scales, to
provide the modified free energy and domain size, and hence the scaling and
criticality.  This description is generalised to low temperatures in
section~\ref{sec:lot-rough} and compared with the numerical domain size and
roughening data, including exponents, from the ground state algorithm in
section~\ref{sec:numrough}.  Section~\ref{sec:final-conc} states the main
conclusions of the paper.

\section{Flat Wall}
\subsection{Zero Temperature Theory}
\label{sec:fwtheory}

\subsubsection{Introduction}

In this section, we establish the domain wall description of the RFIM by taking
the zero temperature case and assuming flat walls.  This is the springboard for
all the further analytic work.  Surprisingly perhaps, the results of even this
very simply case are very good in certain ranges of dimensionality, but they
indicate also where a more comprehensive picture involving roughening is
required, and this is provided in section~\ref{sec:rough}.

The procedure introduced here is to obtain the dispositions of walls by
minimizing an energy which is discussed below.  The resulting domain size is
then used in a phenomenological (finite size) scaling scheme to provide the RG
transformation and hence the critical properties.  Both these steps are most
easily carried out in the bar geometry discussed below.

\subsubsection{The Domain Picture}
\label{sec:dompic}

We suppose, following the prescription of Imry and Ma~\cite{imr75},  that the
contribution of the field to the energy of a domain goes like $h \sqrt{V}$
where V is the number of spins enclosed within a domain and $h$ is the standard
deviation of the field distribution $P$, assumed to have zero mean.  This is
plausible for any distribution of fields that tends towards a Gaussian
distribution in the sense of the central limit theorem.  Also, the energy
change per domain due to the exchange interaction is exactly proportional to
the perimeter of the domain $A$; so
\begin{equation}
\label{eqn:ham}
{\cal F}(T=0) = n_d (2 J A - c_0 h \sqrt{V})
\end{equation}
where $n_d$ is the number of domains.   In this equation $A$ and $V$ represent
an average perimeter and volume measurement for all the domains.  As it is a
constant, the basic ferromagnetic energy $-2J N q$ (where $q$ is the
coordination number of the lattice) is neglected here and throughout this
paper.  The constant $c_0$ is of order unity and represents both the
statistical fluctuations and the selection effects of the domains.  We describe
the origins of this constant, and determine its value, in
section~\ref{sec:TMLoT}.

We have chosen the strip geometry  in which to apply the zero temperature free
energy~(\ref{eqn:ham}) for the following reasons: $(i)$ convenience for
application of finite-size scaling procedures, $(ii)$ amenability to domain
wall arguments, and $(iii)$ we wish to later make contact with transfer matrix
calculations.  With these considerations in mind, we choose a $d$ dimensional
cubic system of finite extent in $d-1$ directions and infinite in the remaining
dimension.
Our analysis gives a length scale $\Xi_L(h)$ which diverges like the
correlation length $\xi_L(h)$ at the fixed point $h^\star$.  We can then
determine the bulk critical properties via an RG equation of the form
 $h \rightarrow h^\prime = R(h)$ which arises from the phenomenological
finite--size scaling ansatz
\begin{equation}
\label{eqn:pfss2}
\frac{\xi_L(R(h))}{L} = \frac{\xi_{bL}(h)}{bL}.
\end{equation}
Standard RG procedures provide from $R_b(h)$ the critical condition and
exponents (from the fixed points and the eigenvalues, respectively).

\subsubsection{The Zeroth Order Theory}
\label{sec:zero}

Given the strip geometry described in the preceding section, the leading order
modification of the ferromagnetic ground state where all the spins are aligned
is a splitting of the system into $n_d$ domains.  These domains are taken to
have flat walls that span the width $L$ of the strip and are of typical length
$\Xi_L$, as shown in Figure~\ref{fig:stripgeom}.  For this picture to be
consistent we need $\Xi_L \gg L$, which will turn out to hold if $h$ is
sufficiently small.  In this calculation, and in the rest of this paper, we set
our units so that the lattice constant is unity.  Using the example of
equation~(\ref{eqn:ham}) we may write the excess energy of this domain state as
\begin{equation}
\label{eqn:flatfe}
{\cal F}(T=0) = n_d \left( 2 J L^{d-1} - c_0 h \sqrt{\Xi_L L^{d-1} } \right)
\end{equation}
where $n_d = N/\Xi_L$.  At zero temperature, this energy is equal to the
Helmholtz free energy of the system.  So, to find the equilibrium free
energy we extremalize with respect to $\Xi_L$ yielding the solution
\begin{equation}
\label{eqn:xiflat}
\Xi_L = L^{d-1} \left( \frac{c_0 h}{4J}\right) ^{-2} .
\end{equation}

Applying the phenomenological scaling equation~(\ref{eqn:pfss2}) gives an RG
equation for $h$ of the form
\begin{equation}
\label{eqn:zero-hrg}
h^\prime = h b^\frac{2-d}{2}
.\end{equation}
This implies that there is a fixed point at $h=0$ which is unstable
for $d < 2$ and stable for $d > 2$.  The canonical RG prescription
combined with~(\ref{eqn:zero-hrg}) gives the critical exponent $\nu =
2/(2-d)$ for the unstable fixed point at $h=0$.  This result holds for all
$d<2$.  However, it also warns us that $d=2$ is a special case where the
scaling field $h$ is marginal.  This is connected with the fact that the lower
critical dimension of the RFIM is $d_{lc}=2$. Higher order terms in the free
energy, however, can break this marginality.  The wall roughening decorations
described in section~\ref{sec:rough} achieve this.

\subsubsection{Relation of the Domain Size to Correlation Length}
\label{sec:domcorr}

In the previous section we introduced the average
domain size, $\Xi_L$.  This is related, as we now show, to the correlation
length $\xi_L$ of the flat wall model.

Consider the configurationally averaged correlation function $G(r)$ between
spins in two columns separated by $r$ lattice units.  Suppose that the flat
domain walls are independently randomly distributed, so that $p$ is the
probability that a domain wall lies between two arbitrarily chosen adjacent
columns.  Then, since all the spins in a given domain are aligned with each
other,
\begin{eqnarray}
G(r) &=& (1-p) G(r-1) - p G(r-1)\nonumber\\
 &=& (1-2p) G(r-1)
\end{eqnarray}
and $G(0)=1$.  It follows that
\begin{equation}
 G(r)=(1-2p)^r \equiv e^{-r/\xi_L},
\end{equation}
where the last relation defines the correlation length $\xi_L$.

However the definition of $p$ shows that its inverse is the average domain size
$\Xi_L$:
\begin{equation}
p = \Xi_L^{-1}
\end{equation}

So we conclude that the correlation length is related to the domain size by
\begin{equation}
\label{eqn:domcorrlat}
\xi_L = \left[ \ln \left( \frac{1}{1 - \frac{2}{\Xi_L} } \right) \right]^{-1}.
\end{equation}
In the limit of large $\Xi_L$ this gives $\xi_L = \Xi_L/2$, which is the same
as provided directly by a continuum version of the analysis given above.

\subsubsection{The Correlation Length at $T=0$}
\label{sec:rw}

If we assume a flat wall picture, then we can actually find the zero
temperature correlation
length of the RFIM on a strip exactly, without employing the more intuitive
energetic arguments presented above.  Using this result, we can check the
results of both section~\ref{sec:zero} and section~\ref{sec:domcorr}.  This
calculation generalizes the work
done by Farhi and Guttmann~\cite{far93} for the one dimensional RFIM.

The key is to consider the connected correlation function
\begin{equation}
\chi_{il} = \langle \sigma_i \sigma_{i+l} \rangle - \langle \sigma_i \rangle
\langle \sigma_{i+l} \rangle
\end{equation}
where we use angled brackets to denote the usual thermodynamic average and
$\sigma_i$ and $\sigma_{i+l}$ are any two spins in columns $i$ and $i+l$,
respectively.
We remark that when $T=0$ the thermodynamic average becomes an average over the
possible ground states of the system.
The trick to finding the dominant behaviour of this function arises from
a desire to devise an algorithm that will solve for the ground state of the
RFIM.  In considering the ground state, one notes that even if the field is
small, the energy to create a domain wall, $2JL$,  can be accumulated from
fluctuations in the random field over large domains.  This is the reason why
there is no long--range order for finite $h$, even at zero temperature, in two
dimensions.  Note that the work of Imry and Ma\cite{imr75} demonstrates that in
$d \le 2$ the $L$ dependencies are such that this result persists in the $L
\rightarrow \infty$ limit.

With this in mind, let us define the quantity
\begin{equation}
{\cal S}(k,l) = \sum_{j=k}^{k+l} \sum_{i=1}^{L} h_{ij}
\end{equation}
where we temporarily adopt the notation that $i$ labels the vertical coordinate
on the strip, and $j$ the horizontal one.
Then, whenever $|{\cal S}(k,l)| \ge 4JL$, it is favourable to insert a domain
wall at positions $k$ and $k+l$.  When we attempt to employ this
algorithm on a variety of field configurations, one immediately notices that
the ground state is still not completely determined.

A region between two opposite random fields, and with endpoints $k,j$ such that
$S(k,j)=0$ (and $j$ is the site closest to $k$ when this is so) is a ``floppy
domain'' (FD) in the sense that
in an average over the ground states, the spins in these domains may
point up or down, and are not fixed like the spins in other regions (``rigid
domains'' (RD)).

One also notices that there is increased degeneracy if $2JL/h$ is an integer.
This is related to the fact that the conditions for both $\uparrow \downarrow$
and $\downarrow \uparrow$ FD can be simultaneously satisfied.
To avoid the additional difficulties involved in the analysis of this
case, we shall assume that $2JL/h$ is not an integer.

It is the degeneracy in the ground state that gives rise to non--vanishing
connected correlations.  Indeed, it is the FD's that contribute exclusively to
$\chi_{il}$.  For instance, if either spin $\sigma_i$ or spin $\sigma_{i+l}$ is
in a RD then $\langle \sigma_i \sigma_{i+l} \rangle = \langle \sigma_i \rangle
\langle \sigma_{i+l} \rangle$.  Furthermore, if
the two spins are in different FD, then they are uncorrelated random
variables, and their contribution to $\chi_{il}$ also vanishes.  Thus, one need
only consider spins that are in the same FD.

Thus, we are left with a combination of three effects to consider
\begin{itemize}
\item The probability ${\cal W}_1(\tilde{l})$ that there exists a FD of size
$\tilde{l} > l$.
\item Given that spin $\sigma_i$ is in a FD, the probability ${\cal W}_2$ that
the spin $\sigma_{i+l}$ is also in the FD.
\item The thermal average and the average over random field configurations
subject to the
constraint that the spins $\sigma_i$ and $\sigma_{i+l}$ are both in the same
FD.  We label this probability ${\cal W}_3$.
\end{itemize}
We also introduce the variable
$$
{\cal L} = \left\lfloor \frac{ 2JL}{h} \right\rfloor +1
$$
where $\lfloor x\rfloor$ denotes the integer part of the non--integer $x$.

We can view the random field configuration on a strip of width $L$ as a random
walk in $1+1$ dimensions.  At
each site along the horizontal direction of the strip, the walk changes
by a height equal to the columnar sum of random field values at that point.
For instance, if we have a binary distribution of fields, then for $L=3$
the height changes by $ \pm 3, \pm 1$, {\it etc}.  Using this view, ${\cal
S}(0,l)$ represents the height of the random walk at $x=l$ (see Figure
\ref{fig:rw}).

If the aggregated change in height of the random walk ever reaches ${\cal L}$
then
we have accumulated enough fields in either the up or down direction to form a
domain.  If there are regions between these peaks where the change in height is
zero, then these are FD.  Figure~\ref{fig:rw} should also make clear the added
ambiguity that arises if $2JL/h$ is an integer.

This random walk picture is very useful in evaluating the three previously
introduced probabilities.  For instance, ${\cal W}_1(\tilde l)$  is the
probability that the random walk returns to the origin after taking $\tilde{l}$
steps.  This is known to fall off exponentially with $l$,
\begin{equation}
\label{eqn:wapprox}
{\cal W}_1(l) \sim \exp \left[ \frac{-l}{\xi_L\left( {\cal L} \right)} \right].
\end{equation}
where $\xi_L({\cal L})$ is a characteristic decay length.
We shall return to an exact determination of this in a moment.

{}From simple
geometrical considerations, ${\cal W}_2 \sim 1/l$.  For the third probability,
we note that
the average over ground states corresponds to an average over domain
wall positions.  Since the number of places for a domain wall to be
inserted between two spins separated by a distance $l$ goes like $l$, we
could estimate the conditional probability ${\cal W}_3$ as $\sim 1/l^2$.  These
are only rough estimates, the probabilities
depend also upon the ratio $h/J$.  But we only need here
 the $l$ dependence, and the important feature of ${\cal W}_2$ and ${\cal W}_3$
is their power law decay in $l$.  We conclude that the only exponential
contribution is from ${\cal W}_1(l)$, and this gives the exponential decay of
$\chi$, so then the $\xi_L({\cal L})$ defined by~(\ref{eqn:wapprox}) is
actually the correlation length.

Now ${\cal W}_1(\tilde{l})$ is  the probability that a directed random walker
in $1+1$ dimensions returns to the origin after $\tilde{l}$ steps, without
hitting  walls at $\pm {\cal L}$.  We solve this problem by reference to the
transfer matrix formalism of random walks~\cite{pri89}.  This involves writing
down a transfer matrix $T(n,m)$ with $n,m \in \{1,2,\ldots,{\cal L}\}$ which
represents the probability that the random walker makes a step from position
$m$ to $n$.  In terms of the RFIM, this means that
\begin{equation}
\label{eqn:tmdefn}
T(n,m) = \mbox{Prob}( \sum_{i=1}^L h_{ij} = (n-m) h ).
\end{equation}
The leading eigenvalue of the transfer matrix, $\lambda({\cal L})$ is related
to the survival probability by
\begin{equation}
{\cal W}_1 (l) \sim \lambda({\cal L})^l
\end{equation}
so that the correlation length of equation~(\ref{eqn:wapprox}) is
\begin{equation}
\label{eqn:xilambda}
\xi_L( {\cal L}) = \frac{-1}{\ln \lambda( {\cal L} )}.
\end{equation}

For $L=1$, the transfer matrix is (using~(\ref{eqn:tmdefn})) tridiagonal and
symmetric, with zeros on the diagonal and $1/2$ on the neighbouring off
diagonals.
We can generalize this result for all $L$, by noting a general pattern in
a {\em row} through the middle of the transfer matrix:
$$
\begin{array}{l|cccccccccc}
(i-j)& \ldots  -4      &    -3    &   -2     &   -1     &   0       &   +1
&   +2      &   +3  &   +4    \ldots\\
\hline \hline
L=1 & \ldots 0		  & 0           & 0           & \frac{1}{2} & 0	          &
\frac{1}{2}  & 0 	        & 0        & 0  	\ldots\\
\hline
L=2 & \ldots 0		  & 0           & \frac{1}{4} & 0           & \frac{2}{4} & 0
         & \frac{1}{4}  & 0        & 0 	        \ldots\\
\hline
L=3 & \ldots 0		  & \frac{1}{8} & 0	      & \frac{3}{8} & 0	          &
\frac{3}{8}  & 0 	        & \frac{1}{8}        & 0 	         \ldots\\
\hline
L=4 & \ldots \frac{1}{16}       & 0 	        & \frac{4}{16}& 0	    &
\frac{6}{16}& 0 	         & \frac{4}{16} & 0        & \frac{1}{16}\ldots\\
\vdots &		  &   	        &   	      &             &             &
&              &          &             \\
\end{array}
$$

The numerators of the elements follow the same pattern as the numbers in
Pascal's triangle, and the denominators are simply $2^L$. This is due to the
Gaussian nature of the random walk whereby $T_L(n,m) = [T_1(n,m)]^L$.  These
matrices are all diagonalized by the same type of vectors:
\begin{equation}
\Phi_j = \sin\left( j \pi q\right).
\end{equation}
where the wavenumber $q$ is chosen to be consistent with the boundary
conditions of the matrix, {\it i.e.} $q=n/({\cal L} + 1)$.  The pattern of
entries
in the transfer matrix gives the eigenvalues of these vectors as
\begin{equation}
\label{eqn:lambda}
\lambda =  \cos^L (q \pi)
.\end{equation}
Clearly, the largest eigenvalue is the one with $q = q_1 = 1/({\cal L} + 1)$.
If $h$ is small, then ${\cal L}$ is large and $q_1$ is small so that
equation~(\ref{eqn:xilambda}) gives the correlation length as
\begin{eqnarray}
\xi_L &=& \frac{1}{L \ln \cos\left(\frac{\pi}{2J/h + 2}\right)} \nonumber \\
&=& 2 L \left( \frac{2 J}{\pi h} \right)^2 + \frac{16 \pi^2}{h} + {\cal O}(1)
\label{eqn:rwxi}
.\end{eqnarray}
It is easy to generalize these
arguments to higher dimensions via the
replacement $L \rightarrow L^{d-1}$.  Then~(\ref{eqn:rwxi}) agrees to highest
order in $h$ with our earlier result for the $T=0$ small $h$ correlation length
(from combining~(\ref{eqn:xiflat}) with~(\ref{eqn:domcorrlat})).  However, it
is important to note that this result came from an analysis of the {\em
connected} correlation function, whereas our prior analysis applied to the {\em
disconnected} correlation function.  Although the correlation lengths derived
from these two quantities should diverge in the same manner, they may differ by
a multiplicative constant ({\it e.g.} $c_0$).  Indeed, this turns out to be the
case in section~\ref{sec:fwtest}.

Although this calculation also assumes a flat wall scenario, it should
be accurate when $\xi_L \gg L$ and is exact for $L=1$.

\subsubsection{Conclusions}
\label{sec:fwt-conc}

In this section,  finite-size scaling in the strip geometry has been used to
study the zero temperature fixed point of the RFIM.   We found the critical
exponent $\nu=2/(2-d)$ for $d<2$.  This result indicates the need for a more
sophisticated analysis when $d=2$, and this  follows in the subsequent
chapters.

The procedure involved the determination of the free energy and the typical
domain size and its relationship to the correlation length.   These results are
confirmed by  a direct calculation of the correlation length of the RFIM on a
strip, assuming flat domain walls, which agrees perfectly  up to the
coefficient $c_0$.

This provides a consistent treatment of the RFIM at zero temperature and low
random field strength away from the lower critical dimension.  However, when
$d=2$ fluctuations are very important to the critical behaviour.  So, the next
section studies the effects of  fluctuations in the domain wall shape, and this
will complete the description of the zero temperature RFIM near the critical
point.  Comparisons with numerical work and investigation of thermal effects
will be provided in later sections.

\subsection{Low Temperature Theory}
\label{sec:lot}

The extension of the domain wall analysis to non--zero temperature requires the
construction of an entropy.  The procedure will then be to minimize the free
energy ${\cal F} = U - T S$, and so determine the characteristic lengths for
exploitation via phenomenological scaling.  By working in the low field and low
temperature limit, {\it i.e.} low $H$ and $T$, we can use
the well separated strip--spanning domain pictures developed in the previous
section.  For the zeroth order (flat wall) description, the entropy then turns
out to be quite trivial, corresponding to the number of ways of laying down
flat
domain walls with an average spacing defined by $\Xi_L = N/n_d$.  Thus, the
entropy is
\begin{equation}
S_0(\Xi_L) = \ln \left( \begin{array}{c} N \\ n_d \end{array} \right) = N
\left[\frac{\ln(\Xi_L -1)}{\Xi_L} - \ln\left(1-\Xi_L^{-1}\right)\right]
\end{equation}
choosing units so that $k_B =1 $ and using Stirling's approximation in the
thermodynamic limit. The free energy generalizing~(\ref{eqn:ham}) to $T \neq 0$
is then
\begin{equation}
\label{eqn:lotfe}
{\cal F} = n_d \left( 2 J L^{d-1} - c_0 h \sqrt{\Xi_L L^{d-1} } \right) - T
S_0(\Xi_L)
.\end{equation}
Extremalizing with respect to
$\Xi_L$ gives the following equation for the domain size
\begin{equation}
\label{eqn:lotxi}
 - 2 J L^{d-1} + \frac{c_0 h}{2} L^{\frac{d-1}{2}} \Xi_L^{1/2} + T \ln
(\Xi_L -1) = 0.
\end{equation}
The various limits of this equation are
\begin{equation}
\begin{array}{lr}
\Xi_L =  \left(\frac{4J}{c_0 h}\right)^2 L^{d-1}, &  T=0 \\
\Xi_L =  1 + \exp\left(\frac{2 J L^{d-1}}{T}\right), &  h = 0  .
\label{eqn:xih0}
\end{array}
\end{equation}
As a check on equation~(\ref{eqn:xih0}), we can use it to investigate the low
temperature
behaviour of the pure Ising model (zero field).  By using this form
for $\Xi_L$ in the phenomenological renormalization equation~(\ref{eqn:pfss2})
 we get the following RG equation for $T/J$
\begin{equation}
\label{eqn:pure-trg}
\left(\frac{T}{J}\right)^\prime = \frac{ 2 L^{d-1} \left(\frac{T}{J}\right) }{2
(bL)^{d-1}  - (\ln b)\left(\frac{T}{J}\right) }.
\end{equation}
This has a fixed point at $T=0$ which is unstable for $d\le1 $; it is stable
for $d>1$, implying the existence of a finite critical temperature which we
know the pure Ising model should have in $d > 1$.  In $d=1$,~(\ref{eqn:xih0})
gives the usual pure Ising result $\xi \sim \exp(2J/T)$.

Thus, in $d=1$ the two scaling variables (scaling like a length) are
$\exp(2J/T)$ and (from~(\ref{eqn:zero-hrg})) $h^{-2/(2-d)}$.  It follows that
in $d=1$ the correlation length can be written in the crossover form
\begin{equation}
\label{eqn:xover}
  \xi = e^{2J/T} \phi(h^2 e^{2J/T})
\end{equation}
where $\phi(x) \rightarrow c_a$ or $c_b/x$ for $x\rightarrow 0$ or $\infty$,
respectively, with $c_a,c_b$ constants.

For $d>1$ the stability with respect to temperature of the $(T,h)=(0,0)$ fixed
point means that no crossover form then applies at low $T,h$, but instead
$h^{2/(2-d)} T^{1/(d-1)}$ is invariant under scaling.

Because of the attractive thermal nature of this $T=0$ fixed point at small $h$
and $d>1$, the thermal scaling has little influence on the low $T$ low $h$
critical behaviour for $d >1$.  Nevertheless, thermal effects contained
in~(\ref{eqn:lotfe}) and~(\ref{eqn:lotxi}) can substantially modify the
behaviour of the free energy and domain size in the low $T$, low $h$ regime for
$d>1$, masking the field dependences more significant for the RG scaling in
this regime.  So~(\ref{eqn:lotfe}) and~(\ref{eqn:lotxi}) will be required for
interpretations in the next section, where numerical techniques are used to
confirm the basic scaling picture so far discussed.

\subsection{Numerical Evaluation}
\label{sec:fwtest}

\subsubsection{Introduction}

We wish to test the results developed in the previous section
by comparing them to independent numerical investigations.  If we find
reasonable
agreement between the two, then one can feel confident
about the veracity of both results.

The structure of the analytical approach is such that the numerical tests can
be applied at various stages.  The most obvious test of such a theory is on its
final scaling predictions for the bulk system criticality.  However, this is
difficult since it requires the investigation of a system large enough to show
the bulk critical behaviour.  It is much easier to apply the numerical
techniques to the finite size strip geometry, as is done here, and to
investigate the non--critical ingredients from which the scaling transformation
is built, and hence establish the applicability of the basic flat wall
procedure.

Three techniques have been found to be successful, and their results are
reported in this paper.  There are a transfer matrix calculation, a modified
Monte Carlo simulation, and a ground state algorithm.

The transfer matrix method is appropriate to the strip geometry, and can in
principle provide the phenomenological RG transformations directly and very
precisely, as has been exemplified for many low dimensional lattice systems,
particularly those with non--random transfer matrices.

Very accurate results have also been obtained for random bond and site diluted
Ising models and a modification of those techniques is used here~\cite{que92}.
Because of the quenched disorder present in such models (and in the RFIM), the
first (dominant) Lyapunov exponent of the transfer matrix product gives the
average free energy of the strip, which is one of the two key ingredients in
our domain wall analysis which we would like to check by the transfer matrix
and other techniques.  The other ingredient is th domain wall size or
correlation length.  Typically, the correlation length is provided by the TM
approach via
\begin{equation}
\xi_L = \frac{-1}{\ln \left(\lambda_1/\lambda_0\right)}
\end{equation}
where $\lambda_0$ and $\lambda_1$ are the largest and second largest
eigenvalues, respectively.  However, this relation gives the {\em most
probable} correlation decay \cite{tm17}, which need not be the same as the {\em
average} correlation decay.  In fact, the correlations may
follow quite complex distribution functions \cite{tm17}, and the difference
between ``most probable'' and ``average'' decays becomes especially important
when the effective interactions may differ in sign, such as in spin glasses or
random field problems \cite{tm17,tm18}.  For this reason, the comparison with
transfer matrix data will be restricted to the free energy.

Monte Carlo simulation is the second technique used.  The usual equilibriation
problems set by critical slowing down are here  compounded by the effects of
very large energy barriers in situations ($d\le 2$) where the criticality is at
$T=0$.  The problems are partly alleviated by the strip geometry, particularly
in $d=2$, where characteristic lengths of order $h^{-2}$
(see~(\ref{eqn:xiflat})) replace exponentially large correlation lengths of the
bulk system (see~(\ref{eqn:xibulk})).  However the equilibriation times remain
excessively long, and a new algorithm (described in more detail in~\cite{mymc})
was devised and employed to markedly improve thermalization times.  Briefly,
this algorithm continues the ideas of multigrid methods with self adjusting
renormalisations of parameters to allow for changes to be made on all length
scales via moves of blocked spins.  This permits the calculation of average
energies, domain sizes, and correlation lengths, for comparison with the
analytic predictions.

The final numerical technique used was an exact algorithm~\cite{mitreport} to
construct the ground states for both binary and Gaussian random field
configurations.  Ground states for models with arbitrary random fields and
arbitrary but not frustrated exchange interactions may be found in polynomial
time by mapping the optimization problem to a min--weighted--cut problem on an
associated graph~\cite{mapping}.  A new min--cut max--flow
algorithm~\cite{mitreport} was first implemented by Ogielski~\cite{ogi86a} to
demonstrate the practicability of this method.  We here apply the algorithm to
the strip geometry, and hereafter refer to this method as the ``max--flow
algorithm''.  From these data it was possible to obtain $T=0$ properties
including average energy, domain size, as well as domain wall roughening
characteristics which will be discussed later.

We now proceed to the numerical data which  test the flat wall picture
discussed so far.  The transfer matrix determinations of the free energy are
first described and compared with the analytic work, followed by the Monte
Carlo results and comparisons, then max--flow results and comparisons.

\subsubsection{Transfer Matrix}
\label{sec:TMLoT}

The simulations were done on strips of width $L=2,\dots,7$ and
length $N=10^5$ and a binary random field distribution.  Periodic boundary
conditions were applied
in the ``finite'' direction whereas all possible trial state vectors
were applied to termini of the strip at each end of the ``infinite''
direction.\footnote{As with all the numerical data presented in this paper,
these calculations were performed on DEC Alpha workstations.}  This procedure
was then repeated for three different realizations of the random field
configuration, and the resulting free energy
\begin{equation}
{\cal F}/N +2 = - k T \ln \lambda_0
\end{equation}
averaged over these realizations and over the $2^L$ trial vectors.  The average
was calculated in the root-mean-square manner, and the associated error in the
free energy was calculated as the rms error in the free energy
at each random field configuration.

We  wish to test the predictions of the zeroth order flat wall theory
(section~\ref{sec:fwtheory}), in which the free energy at very low temperature
and $d=2$ is
\begin{equation}
\label{eqn:TMfe}
{\cal F}_0 =  - \frac{N c_0^2 h^2}{8LJ}
.\end{equation}
To this end,
we have generated data for $L \in (2,7)$ at $T=0.1J$.  For lower temperatures,
the elements of the transfer matrix become larger
than the computer can handle.  If the strip width is much larger than
$7$ the run times become unreasonably long (over a week).  Nevertheless,
these strip widths are adequate for our purposes.

As a test of
this functional form, we plot $\ln({\cal F}_{TM}/J )$ versus $\ln(h/J)$ for
various strip widths $L$, where ${\cal F}_{TM}$ represents the free
energy determined by the transfer matrix method.  Straight lines of slope $2$
are expected,  and this is indeed the case, as demonstrated in
Figure~\ref{fig:TMfefit}.

Straight line
fits to these plots yield the slopes given in Table~\ref{table:Q}.
We include the value of the $Q$ statistic, which is a measure of the
probability
that a value of $\chi^2$ as {\em poor} as the value we have found should
occur by chance.  $Q$ is always in the range $(0,1)$ and a value of $Q > 0.1$
usually denotes a reliable fit~\cite{numrec}.

The agreement with the predicted slope of $2$ is acceptable, with
increasing accuracy at larger $L$.  However, there appear to be systematic
errors, as one can see from a close inspection of Figure~\ref{fig:TMfefit}.
This is an indication of the importance of higher order decoration effects and
finite--temperature effects, occurring even at these low temperatures and high
fields.

We now attempt to obtain a best value
for the constant $c_0$, so that we can compare the predicted free energy with
the numerics, and extract subdominant features noted above.

A simple fit of the numerical free energy to equation~(\ref{eqn:TMfe}) is
statistically unreliable, and gives residual $L$ dependence in $c_0$, and
other artifacts.  We concluded that this was due to the data being at $T=0.1J$
and not zero temperature.  Indeed, for small $h$, the leading temperature
dependence can be very significant.  The zeroth order finite temperature theory
of~(\ref{eqn:lotfe}) and~(\ref{eqn:lotxi}) implies that for large $\Xi_L$ in a
two--dimensional lattice
\begin{equation}
{\cal F} = -N \frac{c_0^2 h^2}{8 L J} \left[ 1 + {\cal O} \left( \frac{T}{L J}
\ln\left[L\left(\frac{4 J}{c_0 h}\right)^2 \right] \right) \right]
.\end{equation}
Since the logarithm diverges for small $h$, the temperature dependence
 becomes increasingly important in this regime.

To fit the free energy data properly, we need a free energy form applying for
all $h,T$ as long as $\Xi_L$ is large.  This can be obtained by
reparameterizing the relations~(\ref{eqn:lotfe}) and~(\ref{eqn:lotxi}) giving
the finite temperature flat wall theory, provided the theory is correct for
large $\Xi_L$, as we have argued.  Using $h,T$ small, $\Xi_L$ large, $d=2$, the
approximate forms to reparameterise are the excess free energy
\begin{equation}
\frac{{\cal F}}{N L} \equiv f = \frac{- c_0 h}{2 \sqrt{ \Xi_L L}},
\end{equation}
and the extremalizing equation
\begin{equation}
\label{eqn:extreme}
\left(\frac{c_o h L^{1/2} }{4T}\right) \Xi_L^{1/2} + \frac{1}{2} \ln \left(
\Xi_L \exp\left(\frac{-2JL}{T}\right) \right) = 0
.\end{equation}
 We now define a new variable $y =  \tilde{c}(T) \sqrt{\Xi_L}$, where
$\tilde{c}(T) \equiv \exp(- L J/ T)$.  Then, the intensive free energy is
\begin{equation}
\label{eqn:TMfebeta}
f = \frac{- c_0^2 h^2}{8 \tilde{\beta} y T }
\end{equation}
and the minimizing equation for $\Xi_L$ has the simple form
\begin{equation}
\label{eqn:TMysc}
\ln y = - \tilde{\beta} y
\end{equation}
where
\begin{equation}
\label{eqn:TMbetadefn}
\tilde{\beta} = \frac{c_0 h \sqrt{L}}{4 \tilde{c} T}
.\end{equation}

We may then eliminate $y$ between~(\ref{eqn:TMfebeta}) and~(\ref{eqn:TMysc}) to
arrive at an equation for $f$ involving $\tilde{\beta}$.  Inserting its
explicit form leads to
\begin{equation}
\label{eqn:TMfeimp}
\frac{L J}{T} = \ln\left( \frac{- c_0 h}{2 f \sqrt{L}}\right) - \frac{c_0^2
h^2}{8 f T}
.\end{equation}
This is an implicit equation for $f$ as a function of $h,T,L$.  We will use it
to solve for $f$ numerically and so determine $c_0$.  The usual $\chi^2$
statistic was calculated
\begin{equation}
\chi^2 = \sum_i \left( \frac{f(c_0 h_i, T_i, L_i) - {\cal F}_{TM}}{\sigma_i}
\right)^2
\end{equation}
where $\sigma_i$ is the estimated error on ${\cal F}_{TM}$.  This statistic was
minimized by a bisection routine\footnote{The more efficient Newton-Raphson
type routine could only be used if we had numerical derivatives of $f$.  These
are not easily found from equation~(\ref{eqn:TMfeimp}).}, which also calculates
the $Q$ statistic.

As well as obtaining $c_0$, we can test whether non--Gaussian effects occur for
small $L$, or perhaps the
flat wall description breaks down for large $L$ by varying
the range of the fit.  The values of $T$ used were $0.1 J, 0.2 J$ and $ 0.3 J$,
with $h \in \{0.05,0.25\}$.  Over this range of $T$ and $h$, the zeroth order
theory predicts large correlation lengths at least $> 10 L$ for even
a liberal estimate of $c_0$, which should make~(\ref{eqn:TMfeimp}) applicable.
This assumption was also demonstrated to be
self--consistently true by using the fitted value of the selection constant to
predict $\Xi_L$.  Over this range of fields, then, the results are presented in
Table~\ref{table:Q2}.

One notices the striking improvement to the fits when the lower $L$ values
are dropped ($Q=0.686$ is remarkably good).  We attribute this to non--Gaussian
selection effects at
lower $L$ values, where the domains do not sample over as many field
configurations as a domain of the same size at large $L$.  Since the number of
field configurations for a column of length $L$ is $2^L$, we would expect
a marked decrease in this effect as we increase $L$.

The value $c_0 \approx 1.85$, found by fits of the theory to the free energy
data, does not agree with that $(c_0=\pi)$ predicted in section~\ref{sec:rw} by
a random walk analysis of the disconnected correlation function.  As explained
in section~\ref{sec:rw}, this is due to the different correlation functions
used in obtaining this (non--universal) constant.

The large $Q$ statistic gives much confidence in the form of ${\cal F}$
predicted from the flat--wall {\em ansatz}.

Figure~\ref{fig:fefit1} illustrates the quality of the agreement between the
predicted
form of the free energy and that predicted by equation~(\ref{eqn:TMfeimp})
using $c_0 = 1.85$.\footnote{The data at $T=0.2J$, while it was used in the
fits, displays no new features.}.  One should remark that, while the fits were
done for $h \in (0.05, 0.25)$, the predicted free energy seems to lie within
error for a much larger regime, especially for larger strip widths.  Indeed,
for $L=8,9$ the predicted free energy lies within error of the transfer matrix
data all the way out to $h=1.0 J$.

Transfer matrix data were also taken for $h/J \in (1,6)$ and low $T$, to
determine
the effect of this increasing field strength.  Typical results are shown in
Figure~\ref{fig:TMhrange}.  One notices that there are three separate
regimes in the graphs:
\begin{itemize}
\item  $h/J \in (0,\approx 2)$.  In this regime, the free energy seems to go
roughly like the square of the field.    As emphasized above, this dependence
is close to that
predicted by the flat wall theory, with increasing accuracy as the strip
width, $L$, increases, which is important for our scaling discussion of
critical effects.
\item $h/J \in (\approx 2,4)$.  Here, the free energy crosses over
from quadratic to  linear dependence on $h$, due to a rapidly decreasing domain
size.
\item $h/J > 4$.  The field strength is greater than the energy cost of
flipping a
spin, and so all the spins align with their local field.  Thus the free energy
is linear in $h$, with unit slope.
\end{itemize}

To sum up: the transfer matrix approach has provided a rigorous test of  the
energetic arguments of section~\ref{sec:zero} and~\ref{sec:lot}.  It provided
excellent agreement over the range tested, giving confidence in the
applicability of the theory.

However, a comparison of quantities like correlation length, domain size, etc.,
would be a more rigorous check and would also provide a better picture of the
physics underpinning the results.  Secure results on these lengths are not so
far available from the transfer matrix method and so we turn to the Monte Carlo
analysis, which can provide them.

\subsubsection{Monte Carlo}
\label{sec:mc}

Using the Block Monte Carlo algorithm described in reference~\cite{mymc} we now
investigate other physical quantities which are directly acessible via our
theoretical framework, but not easily obtained by transfer matrix analysis.

\paragraph{The Average Energy:}
\label{sec:mcave}
First, we compare the configurationally averaged energy found by the
Monte Carlo routine to that predicted by the flat wall results of
chapter~\ref{sec:lot}.  This quantity is the first term in~(\ref{eqn:lotfe})
and involves $\Xi_L$ which is given by numerical solution of~(\ref{eqn:lotxi}).
 The theoretical and Monte Carlo data points are shown for $L=1$ and $4$ in
 Figure~\ref{fig:mcavel14}.

Similar results are obtained from the data for $L=2$. Fitting the data to the
theoretical form~(\ref{eqn:flatfe}) gives
 $c_0 = 1.75 \pm 0.05$, in reasonable agreement with $c_0 = 1.85$ needed to fit
the transfer matrix free energy data (section~\ref{sec:TMLoT}).

The theory agrees quite well with the Monte Carlo data up to a  temperature,
that diminishes with increasing field strength, at which $\Xi_L \approx L$ and
the criterion for the validity of the theory breaks down.  Yet, there is
agreement within error up to $T=0.7 J$ even for $h=0.5 J$ when $L=4$.  This is
a further indication that the zeroth order theory captures the essential
flavour of the RFIM.

\paragraph{The Domain Distribution and Correlation Length--Domain Size
Relationship:}

A key assumption in our derivation of the zeroth order entropy
(section~\ref{sec:lot}) and of a relation between domain size $\Xi_L$ and
correlation length $\xi_L$ (section~\ref{sec:domcorr}) was that the domain
walls were randomly placed on the strip.  This can be tested either by
investigating the domain size distribution, which should then be Poisson, or by
exploring the predicted relation between $\Xi_L$ and $\xi_L$.

Since the standard deviation and mean of a Poisson distribution are the same,
we have evaluated these two quantities for $L=1$, 2 and $4$ for $h \le 0.5J$
and $T \le J$.  In every case they are found to be the same within error (never
more than $5 \%$).  While this result is not definitive, it is very suggestive
of a Poisson distribution.

Equation~(\ref{eqn:domcorrlat}) expresses the relation between the correlation
length $\xi_L$, and the domain size $\Xi_L$ when on a lattice, for randomly
distributed walls.  Figure~\ref{fig:mccorrl24} shows a test of this
relationship for $L=2$.  Very similar plots arise for $L=1$ and $4$.  In each
case, the agreement with the predicted form is remarkable, and gives us
confidence that, at least for large correlation lengths, the methods used in
section~\ref{sec:domcorr} are correct and the domain walls really are Poisson
distributed.

The final support for this conclusion comes from direct analysis of the
probability distribution of domain sizes itself for $L=1,2,4$; $h=0,0.2J$ and
$T$ up to $J$.  The results for $L=1$ are shown in
Figure~\ref{fig:mcdomdistl1}.
The resulting size distribution is very close to exponential (providing strong
evidence that the domains really are Poisson distributed) for $h=0$, at least
for smaller $L$'s and larger temperatures.  However, when $h=0.2J$ and $T=0.4
J$ the distribution seems to decrease at lower $\Xi$.  This is evidence for
{\em domain wall repulsion}, which is expected at higher fields when the domain
walls begin to approach and undergo appreciable roughening (see
section~\ref{sec:rough}).  For $h < J$, however, this effect is only noticeable
over about the lower $2\%$ of the domain size distribution.  Otherwise, the
distribution is still exponential.  So, the Poisson assumption appears to be
approximately correct, and gives the correct behaviour at large domain sizes.
Thus, the flat wall entropy should still be a valid starting point, although a
more complete description is obviously required for a full understanding,
particularly in the large size regime of interest for scaling.

\paragraph{The Domain Size Itself:}

Having justified assumptions implicit in the flat wall theory, we wish to test
its predictions further.  In this section,  the domain size $\Xi_L$, predicted
by~(\ref{eqn:lotxi}) is compared with that measured by the MC routine.

The results are shown for $L=1$ and $4$ in Figure~\ref{fig:mcdoml24}.  The data
for $L=2$ looks similar.  The flattening of the data at low $T$'s is due the
finite length of the simulated system; data in the flattening regime are
discarded in quantitative comparisons.  The constant $c_0$ was determined by a
best fit to the data at $T=0.1 J$, where we might expect the flat wall theory
to be accurate.   The result, $c_0=1$,  contrasts with $c_0 = 1.75 \pm 0.05$
given by the average energy fits presumably because selection effects  act
differently
for length scales and energetic quantities.

Contrary to the previous convincing agreements with theory, the comparison of
the predictions of~(\ref{eqn:lotxi})  to the
domain size data is relatively poor.  The reason for this poor agreement is not
well understood, but may be due to non--Gaussian effects at smaller strip
widths.  However, we do get reasonable agreement (discarding flattened data) at
the lowest of temperatures, and for larger $L$.  These are the conditions that
matter most for the scaling constructions, where the correlation length and
domain size at $T=0$ and large $L$ is of particular importance.  These are
discussed below using data from the ground state algorithm.

\subsubsection{Max--Flow algorithm}

We have used the max--flow algorithm~\cite{ogi86a} to find ground states and
hence ground state properties for many random field configurations generated
from binary or Gaussian distributions.  In the large scale limit, these two
distributions are expected to give the same results, but there are differences,
discussed in section~\ref{sec:numrough}, for small system or domain sizes.
These differences were not as pronounced in the finite $T$ data obtained from
the transfer matrix and Monte Carlo routines described above. In the present
case, it becomes advantageous to use results from the Gaussian distribution as
we now do.

These data were generated on strips of width $L=2,\ldots,11$ and length
$N=10^3$.  These results were averaged over $100$ independent random field
configurations and the error bars shown represent the statistical variations in
the data when averaged over these runs (as opposed to the width of the
distribution of a variable quantity like the domain size).  As the
computational time required to generate a ground state via this algorithm grows
like $n^3$, and the run time for an $n = 11 \times 10^3$ system was on order 10
days, it was not possible to significantly increase the system size beyond this
limit.

The typical domain size $\Xi_L$ is shown in Figure~\ref{fig:gscorr}.  Also
included in the Figure is a comparison with the theoretical $\Xi_L$ obtained
from the random walk analysis (section~\ref{sec:rw}) of the correlation length
$\xi_L$ using a simple fitted constant of proportionality to convert this to a
domain size.  A quantitative assessment of the quality of the agreement is
provided by the correlation coefficient $r=0.9968$ and the statistic $\chi^2 =
0.426$.  The fit gives a value of $c_0 = 1.98 \pm 0.02$, in reasonable
agreement with the values obtained by both the transfer matrix and block Monte
Carlo results at finite $T$.

Also considered was a fit to just the $h^{-2}$ leading term in the small $h$
expansion of the theoretical $\Xi_L$.  This corresponds to~(\ref{eqn:xiflat})
({\it i.e.} to the basic $T=0$ theory and differs from the full $\Xi_L$ by
${\cal O}(1/h)$, significant except in the low $h$ limit.  The comparison of
fits to this reduced form with those to the full $\Xi_L$ show that if fields up
to $h=2.0J$ are included , the fit to the reduced form gives a  $\chi^2=4.215$.
 This shows a much poorer quality of fit than that to the full theoretical
form.  The neglect of the ${\cal O}(1/h)$ terms is one of the sources of the
reduced quality fit observed in the basic theory analysis of the Monte Carlo
domain size data (section~\ref{sec:mc}).  However, it is difficult to
incorporate these higher order terms in a consistent finite temperature theory.
 This shortcoming disappears in the small $h$ regime important for scaling.

This completes the comparisons of numerical analyses of (transfer matrix, block
Monte Carlo, ground state algorithm)  with the basic theory.  The conclusion is
that this theory provides an excellent account of the free energy and average
energy; and an account of the domain size and correlation length which differs
from the numerical  data by effects whose origin is becoming clear,
particularly so at low temperature.  These effects are the domain wall
repulsion due to non flat wall effects (wall roughening), seen in the domain
distribution of section~\ref{sec:mc}, and the (understood) ${\cal O}(1/h)$ and
other terms by which the basic theory differs from the random walk one.  For
these reasons, and for those given in section~\ref{sec:fwt-conc} it is
necessary to extend the theory by including wall roughening effects and to
verify their importance, which is done in the next section.

\section{Domain Wall Roughening}
\label{sec:rough}
\subsection{Theory}
\label{sec:rough-theory}
\subsubsection{Introduction}

The effect of domain wall
roughening on the free energy, and
hence the correlation length is estimated here by a decoration method
applicable in general dimensionality.
Lattice and continuum cases will be studied, and these turn out to have
different roughening characteristics. We first outline a simple yet effective
calculation (given previously in~\cite{intro}), first on a lattice and then in
the continuum.  Then, a functional field theory in the continuum is devised
which provides a quantitative description of the effects of wall roughening.
The varying results of these three approaches will then be compared.  These
results are next used to perform wall roughening on all length scales.  This
provides an RG transformation including the terms which break the marginality
of  $d=2$.  Hence, we can derive the scaling behaviour in $d=2$, and the
critical point and critical exponents of the RFIM in $d=2+\epsilon$.  All these
calculations are carried out at zero temperature for simplicity, and will be
later generalized to finite temperature.

\subsubsection{The Lattice Theory}

On a lattice, the basic perturbation of the domain wall (``decoration''), to be
later introduced on all scales, can be taken to be a
rectangular shape (see Figure~\ref{fig:balat}),  of width $a$ and
height $b$.  The associated excess
volume $\delta V$ and the excess area $\delta A$ are, in $d$ dimensions
\begin{equation}
\label{eqn:valat}
\delta V = b a^{d-1}  , \;\;
\delta A = 2 (d-1) b a^{d-2}
.\end{equation}
We assume that the domains are sufficiently large and the domain wall
fluctuations sufficiently small ({\it i.e.} $\Xi_L \gg b$) that the
fluctuations in the wall may be considered as uncorrelated from those in the
body of the domain. Thus, we may treat the decoration as a separate entity from
the bulk of the domain and apply the free energy~(\ref{eqn:ham}) independently
to it.

Substituting the values~(\ref{eqn:valat}) into equation~(\ref{eqn:ham}) gives
the
zero temperature excess free energy
\begin{equation}
\label{eqn:lat-fe}
\delta{\cal F}(T=0) = 4 J (d-1) b a^{d-2} - c_0 h b^{1/2} a^{(d-1)/2}
.\end{equation}
We minimize this
free energy with respect to $b$ to arrive at the relation
\begin{equation}
\label{eqn:blat}
b=\left(\frac{c_0 h}{8 J(d-1)}\right)^2 a^{3-d}
.\end{equation}
This relationship is of the generic form
\begin{equation}
\label{eqn:zetadefn}
b=c (h/J)^\kappa a^\zeta
\end{equation}
which defines a wandering exponent $\zeta$, and a further ``field'' exponent
$\kappa$.  We see that $\zeta = 3-d, \kappa = 2$ in this simple description

Inserting~(\ref{eqn:blat}) into~(\ref{eqn:lat-fe}) gives the
equilibrium value of the excess free energy and in particular its power law
dependence on field and on the scale $a$,
\begin{equation}
\delta{\cal F}(a) = \frac{-c_0^2 h^2}{J} \frac{a^{3-d}}{16(d-1)}
.\end{equation}

\subsubsection{The Simple Ansatz for the Continuum}

We now perform a similar calculation to that in the above section
but in a continuum rather than a lattice.  It turns out that these
two cases have different exponents.

By analogy to the previous section, the basic decoration
 of the domain wall is  described by
 the typical height $b$ and the typical width $a$
(see Figure~\ref{fig:balat}).  These will again be related in the
manner~(\ref{eqn:zetadefn}).
Assuming axial symmetry, we denote the excursion of the domain wall from its
mean position as $z=b f(r/a)$ where $f(0)=1$.  Then, the excess surface area in
the free energy~(\ref{eqn:ham})
is
$$
\begin{array}{lll}
\delta S(a,b) &=& \int_0^a {\rm d} r r^{d-2} \left(\sqrt{1+\left( b
\frac{\partial}{\partial r} f \left(\frac{r}{a}\right)\right)^2} -1\right) \\
   &\rightarrow&
   \left\{ \begin{array}{lr}
     b^2 a^{d-3} &  ,\hspace{0.15in}\mbox{if}\; b \ll a \\
     b a^{d-2} &  ,\hspace{0.15in}\mbox{if}\; b \gg a
    \end{array}\right.
\end{array}
$$
Constants of proportionality have been disregarded here, and will be elsewhere
in this section, since they do not affect the exponents in which we are
interested.

Similarly, the excess volume is
\begin{equation}
\delta V(a,b) = \int_0^a {\rm d} r r^{d-2} b f\left(\frac{r}{a}\right) \sim b
a^{d-1}
.\end{equation}

Using the above forms in the free energy~(\ref{eqn:ham}) and
minimizing with respect to $b$ yields the following relations
\begin{eqnarray}
\label{eqn:ba1}
b \sim &\left(\frac{h}{J}\right)^{2/3} a^\frac{5-d}{3}&
,\hspace{0.15in}\mbox{if}\; b\ll a \\
b \sim &\left(\frac{h}{J}\right)^2 a^{3-d}& ,\hspace{0.15in}\mbox{if}\; b \gg a
\label{eqn:ba2}
.\end{eqnarray}

For self--consistency, we must require that~(\ref{eqn:ba1}) holds at large
scales $a$ only for $d > 2$ and that $h/J$ is not much larger than unity, or
that $d=2$ and $h/J \ll 1.$
Similarly, equation~(\ref{eqn:ba2}) only holds if $d < 2$.  We now check the
assumption that the wall decorations are decoupled from the bulk of the
domains, {\it i.e.} that $b(L) \ll \Xi_L$.  Using the form of $\Xi_L$ given by
the zeroth order decorations~(\ref{eqn:xiflat}) and the solutions for $b(L)$
given in equations~(\ref{eqn:ba1}) and~(\ref{eqn:ba2}) above we conclude that
$h/J$ must satisfy
\begin{equation}
\label{eqn:simple-sc}
\frac{h}{J} \ll \left\{ \begin{array}{lr} L^\frac{d-2}{2} &
,\hspace{0.15in}\mbox{if}\; b \ll a \\
                                          L^\frac{d-2}{2} &
,\hspace{0.15in}\mbox{if}\; b \gg a.
                        \end{array} \right.
\end{equation}
In either case then, for large $L$ and $d\ge2$ there is a generous regime for
$h/J$ which satisfies the assumption.  When $d<2$, if $L$ is finite we can
always
find $h/J$ sufficiently small that~(\ref{eqn:simple-sc}) is satisfied, but the
allowed regime for $h/J$ decreases as $L$ increases.

Thus, these simple arguments give the wandering exponent of the domain wall,
$\zeta$,
\begin{equation}
\zeta = \left\{ \begin{array}{ll} \frac{5-d}{3} & ,\hspace{0.15in}\mbox{if}\; d
\geq 2 \\
                                         3-d & ,\hspace{0.15in}\mbox{if}\; d
\leq 2 \; \mbox{and} \; h \ll 1\\
                   \end{array} \right.
\end{equation}
and the free energy
\begin{equation}
\delta{\cal F} (a) \sim \left\{ \begin{array}{ll} -\frac{h^{4/3}}{J^{1/3}}
a^\frac{d+1}{3} & ,\hspace{0.15in}\mbox{if}\; d \geq 2 \\
                                                          -\frac{h^2}{J} a &
,\hspace{0.15in}\mbox{if}\; d \leq 2 \; \mbox{and} \; h \ll 1.
                                    \end{array} \right.
\end{equation}

\subsubsection{The Full Field Theory}
\label{sec:fullfield}

A more complete treatment of this problem allows the entire shape of the
interface to be determined by the system, and not just the height to width
ratio as in the previous treatment.  This will provide a check on the results
of the previous section, and will provide more detailed information on the
average equilibrium shape of the domain walls, and on proportionality
constants.

In order to achieve this, we will establish a functional equation for the shape
of the interface by rewriting the free energy of the domain wall as a
functional of the wall profile $P$.  The equilibrium wall profile is then found
by extremalizing this free energy via the Euler--Lagrange equation.  The
results
of this section only hold in $d \ge 2$.

We use a $d-1$ dimensional cartesian coordinate system
 placed on the hyperplane where the flat interface would have been.  Then,
the deviation of the domain wall from this hyperplane is given by the function
$P(\vec{x})$.  Using the model of equation~(\ref{eqn:ham}), this decoration has
a free energy at $T=0$ of
\begin{eqnarray}
  {\cal F}[P(\vec{x}),\nabla P(\vec{x})] &=& 2 J \int \sqrt{1+(\nabla P)^2}
{\rm d}^{ d -1} \vec{x} \nonumber \\
                                      & & \mbox{}     -c_0 h \sqrt{\int
P(\vec{x}){\rm d}^{ d -1} \vec{x}}
\label{eqn:fedec}
\end{eqnarray}
where both integrals are over the hyperplane defined by $x_i \in [0,L]$.
This free energy is minimized via the generalized Euler--Lagrange equation:
\begin{equation}
  \sum^{ d -1}_{i=1} \frac{\partial}{\partial x_i} \frac{ \delta {\cal
F}}{\delta (\partial P(\vec{x})/\partial x_i)} = \frac{ \delta {\cal F} }{
\delta P(\vec{x})}
.\end{equation}
Note that  the RHS is a constant, and define it as
\begin{equation}
\label{eqn:conN}
N_0 J = \frac{-c_0 h}{4 \sqrt{\int P(\vec{x}) {\rm d}^{ d -1} \vec{x}} }
.\end{equation}

This leads to a rather complicated  $d$--dimensional differential equation.
\begin{equation}
\label{eqn:compde}
N_0 = \sum_{i=1}^{d-1} \frac{ \left\{ \partial_i^2 P \left[1 + \left( \nabla P
  \right)^2\right] - \sum_{j=1}^{d-1} \partial_i P \, \partial_j P \,
  \partial_{ij} P \right\} }{ \left[ 1+(\nabla P)^2\right]^{3/2} }
.\end{equation}
Here, $\partial_i \equiv \frac{\partial}{\partial x_i}$.

To simplify things we consider the isotropic solution, which only depends on
the radial coordinate $r^2 = \sum_{i=1}^{d-1} x_i^2$.

Then, with $\partial_r P(\vec{x}) \equiv P^\prime(r)$ the isotropic equation
becomes
\begin{equation}
\label{eqn:isode}
N_0 = \frac{ \left(\frac{d-2}{r}\right) P^\prime(r)\left(1 +
P^\prime(r)^2\right)
 + P^{\prime\prime}(r) }{\left(1 + P^\prime(r)^2\right)^{3/2} }
.\end{equation}

Note that this equation may alternatively be obtained by extremalizing an
isotropic free energy functional,
\begin{eqnarray}
\delta{\cal F}[P(r),P^\prime(r)] &=& 2 J \Omega \int \sqrt{1+P^\prime(r)^2}
r^{d-2} {\rm d} r \nonumber \\
& &\mbox{} - c_0 h \sqrt{\Omega \int P(r) r^{d-2} {\rm d} r}
.\end{eqnarray}
Here
\begin{equation}
\label{eqn:omdef}
\Omega = \frac{\pi^\frac{d-1}{2}}{\left(\frac{d-1}{2}\right)!} \left(d-1\right)
\end{equation}
is the area of the $d-1$ dimensional unit ball.  However, in the more general
case the isotropic
solution is a member of the larger set of optimal solutions.

It may be checked by direct substitution that the solution to the wall
decoration problem encapsulated in equations~(\ref{eqn:conN})
and~(\ref{eqn:isode}) is a hypersphere of radius $R$
\begin{equation}
  P(r) = \sqrt{R^2 -r^2} - P_o,
\end{equation}
where $R$ satisfies the requirement ${-1}/R = N_0$.  The constant $P_0$ is
determined by the choice of the boundary conditions $P(r=L/2) = 0$
\begin{equation}
P_0 = \sqrt{R^2 - \left(\frac{L}{2}\right)^2 }
.\end{equation}
In addition, the constraint~(\ref{eqn:conN}) imposes the self--consistency
relation
\begin{equation}
\label{eqn:isosc}
  R^2 = \frac{J^2(d-1)^2 4 (\delta V) }{ c_0^2 h^2 }
\end{equation}
which we use to determine $R(h)$.
Here $\delta V$ is the extra volume added by the decoration
   $\delta V = \Omega \int_0^{L/2} P(r) r^{d-2} {\rm d} r$.

Thus,~(\ref{eqn:isosc}) is difficult
 to solve for $R(h)$.  However,  we are
interested in the regime $h/J < 1$ where we expect the decorations to be
small  $(R \gg L)$ and we may expand in powers of $L/R$.  To first order in
$1/R$ the excess volume and area are
\begin{mathletters}
\label{eqn:dv-ds}
\begin{eqnarray}
\label{eqn:dv1}
\delta V &=& \frac{\Omega}{R(d^2-1)} \left(\frac{L}{2}\right)^{d+1} \\
\label{eqn:ds1}
\delta A &=& \frac{\Omega}{2R^2(d+1)}\left(\frac{L}{2}\right)^{d+1}
.\end{eqnarray}
\end{mathletters}
Then, equation~(\ref{eqn:isosc}) gives the equilibrium radius as
\begin{equation}
\label{eqn:eqrad}
R = \left(\frac{c_0 h}{J}\right)^{-\frac{2}{3}} L^{\frac{d+1}{3}} \left(
\frac{4 \Omega (d-1)}{2^{d+1} (d+1)} \right)^{\frac{1}{3}}
.\end{equation}

Inserting the first order forms~(\ref{eqn:dv-ds}) evaluated
 at the equilibrium radius~(\ref{eqn:eqrad}) into~(\ref{eqn:fedec}) then gives
the first order free energy fluctuation
\begin{equation}
\label{eqn:dFhs}
  {\cal F}_{wall}(L) = - \tilde{c} \left(\frac{c_0 h}{J}\right)^{4/3}
L^{\frac{d+1}{3}} + {\cal O} (h^{8/3})
.\end{equation}
Here $\tilde{c}$ is the numerical constant
\begin{equation}
  \tilde{c} = \left[ \frac{\Omega^{1/3}}{ (d+1)^{1/3} (d-1)^{2/3} }\right]
\left[ \frac{ - 2^{-1/3} + 2^{8/3} }{ 2^{(d+7)/3} }\right]
\end{equation}
and is positive for all $d > 1$.
This shows that wall roughening is always favourable for $d \ge 2$.

Finally, the wall wandering height is
\begin{equation}
b \equiv P(r=0) \sim h^{2/3} L^\frac{5-d}{3} + {\cal O}(h^{8/3})
.\end{equation}

We now compare these three approaches to the wall decoration
problem on the basis of their predicted
behaviours of the wall wandering height $b$ and the free energy of
the wall $\delta{\cal F}$.
We define the wandering exponent $\zeta$, the free energy
fluctuation exponent $\omega$, and associated exponents $\gamma$
and $\kappa$ by the following equations
\begin{mathletters}
\label{eqn:general-forms}
\begin{eqnarray}
\label{eqn:f-general}
  \delta {\cal F} & \sim & h^\gamma a^{\theta} \\
  b           & \sim & h^\kappa a^{\zeta}
\label{eqn:b-general}
,\end{eqnarray}
\end{mathletters}
where $a$ is the base length ($L$ in the field theoretic description).
One sees that all the exponents of the simple continuum ansatz and the full
field
theory agree for $d \ge 2$.  The complete comparison between all three
treatments is given in Table~\ref{table:wallexps}, and the results are shown
graphically in Figure~\ref{fig:expplot}.

The predicted exponents in the continuum model agree with those of other
extensions of the Imry-Ma type argument done by Natterman~\cite{nat88}.  On the
other hand, the lattice exponents agree with the work of Binder~\cite{bin83},
also for the lattice.  Grinstein~\cite{gri84}, finds the same values for
$\gamma$, $\kappa$, and $\zeta$ as we do, but finds $\theta = 1$.  It is also
worth noting that the replica symmetric approach of Parisi and
Sourlas~\cite{par79} for the continuous model gave $\zeta = (5-d)/2$, but the
inclusion of replica symmetry breaking gives $\zeta =
(5-d)/3$~\cite{mez90,mez92}.  Finally, the lattice result is consistent with
the
results of transfer matrix calculations done in two dimensions by Fernandez
{\em et al.}~\cite{fer83} and with numerical ground state calculations reported
in section~\ref{sec:numrough}.

\subsubsection{Decorations upon Decorations}
\label{sec:decondec}

The decoration to the flat domain walls described in
section~\ref{sec:fullfield} is (to lowest order in $h/J$), the most
favorable twice differentiable form of the
domain walls.  However, discontinuities
 may be a feature of the ``true'' solution.  We may
incorporate such discontinuities
by allowing the decoration to be repeated on all length scales,
following a procedure of Binder~\cite{bin83}.  Basically,  having made a basic
change in
the shape of the domain wall on scales of length $L$, one then looks
at the domain
wall on a length scale $L/n$ where $n > 1$ is some arbitrary integer, whose
choice, we hope, should not strongly affect the results.  If the random field
is small,  the
wall will be slowly varying on such a scale and may be viewed as a hyperplane,
or to preserve the simplicity of the discussion by taking $d=2$, as a straight
line.  This straight line is then decorated in the same manner as the original
wall.  This process is repeated on length scales of order $a_i = L/n^i,
i=2,3,\ldots$
 until a cutoff length scale
is reached (see Figure~\ref{fig:decondec}).  Reverting to the case of general
$d$, this cutoff occurs at a stage $k$
where the smallest length scale of decorations is the
lattice constant (unity), {\it i.e.}
\begin{equation}
\label{eqn:minab}
\mbox{min}\;(a_k,b_k) = 1.
\end{equation}
{}From the results of the previous section, we know that $b_k = c_1
(h/J)^\kappa a_k^\zeta$ where $c_1$ is a constant (equal to $(c_0/8(d-1))^2$
for the lattice case).  Now $\kappa$ is greater than zero for both the lattice
and the continuum.  Thus, for small $h$, (\ref{eqn:minab}) becomes $b_k = 1$ so
the final level of decorations, $k$, is determined by
\begin{equation}
\label{eqn:binder-cutoff}
a_k = L/n^k =  c_1^{-1/\zeta} h^{-\kappa/\zeta}
\end{equation}
This gives the cutoff $k$ as
\begin{equation}
k = \frac{ \ln L + \frac{1}{\zeta} \ln c_1 h^\kappa}{\ln n}
.\end{equation}
This result should be approximately valid for large $L$.

Finally, all the changes to the free energy from all the
various decorations are summed to arrive at a total decoration.
At stage $i$ we have decimated the length scale by a factor
of $n^i$ so there are $n^{i(d-1)}$ decorations at this stage, and each gives a
free energy contribution $\delta{\cal F}(L/n^i)$.  Hence, the total free energy
of all the
decorations will be
\begin{equation}
\label{eqn:binder-d}
 \delta{\cal F}^{\rm Tot} (L) = \sum^k_{i=0} n^{i(d-1)} \delta{\cal F}
\left(\frac{L}{n^i}\right)
\end{equation}
with $k$ as above.

In this way, then, we can estimate the effect of wall roughening on all
length scales, and remove the restriction that the wall shape must
be smooth.  Such generalizations are crucial
to get the correct marginality breaking free energy, as we shall see.

\subsubsection{Scaling and criticality in  $d=2$ and $d=2+\epsilon$ }
\label{sec:zerot-eps}

The preceeding results can now be used to construct the RG transformation in
$d=2+\epsilon$.  This allows us to investigate the marginality
breaking effect of domain wall roughening in two dimensions, and to
investigate the behaviour of the $h=0$ fixed point as we move away
from $d=2$.  So we obtain the two dimensional critical behaviour and also an
idea of the critical properties in higher dimensionalities.

Using~(\ref{eqn:binder-d}) and the free energy forms found in the previous
sections,  a  full decoration of the domain walls may be done in general
dimension $d \ge 2$.  For the sake of
generality we will use the general form of the wall free energy given
by~(\ref{eqn:f-general}), though with restored generic constants as well as
exponents, which can be read off from earlier results for specific cases.
  Thus, the general form of the wall free energy is
\begin{equation}
\delta{\cal F}^{\rm Tot} = - \tilde{c} \left(\frac{c_0 h}{J}\right)^\gamma
L^\theta \sum_{i=0}^k n^{i(d-1-\theta)}
.\end{equation}
{}From the values of $\theta$ given in Table~\ref{table:wallexps}, the summand
is $$
n^{i(d-1-\theta)} = \left\{ \begin{array}{lll}
 n^{i(\frac{2d-4}{3})} & = 1+i \frac{2}{3} \epsilon \ln n
&,\hspace{0.1in}\mbox{continuum} \\
 n^{i(2d-4)} &= 1+i 2 \epsilon \ln n &,\hspace{0.1in}\mbox{lattice}
                       \end{array}  \right.
$$
for $d=2+\epsilon$.
To further simplify matters we define the coefficient
\begin{equation}
\phi = \left\{ \begin{array}{ll}
\frac{2}{3} &,\hspace{0.1in}\mbox{continuum} \\
2 & ,\hspace{0.1in}\mbox{lattice} \\
           \end{array}  \right.
\end{equation}

Then, the total free energy of the wall decorations may be concisely
written as
\begin{equation}
\label{eqn:first-dfwt}
\frac{\delta{\cal F}^{\rm Tot}}{J} = -\tilde{c} \left(\frac{c_0
h}{J}\right)^\gamma L^\theta\left[ (k+1) + \epsilon \phi \left(\ln n\right)
\frac{k(k+1)}{2} \right].
\end{equation}
This result may be compared both with that of Villain~\cite{vil82} and
Grinstein and Ma~\cite{gri82} who find $\delta{\cal F}^{\rm Tot} \sim h^{4/3}L
\ln L$ for $d=2$ using a continuum interface model, and with the result of
Binder~\cite{bin83} who finds $\delta{\cal F}^{\rm Tot} \sim h^2 L \ln L$ in a
$2d$ lattice calculation.  Both these results agree with~(\ref{eqn:first-dfwt})
with the appropriate choice of the exponents $\gamma$ and $\theta$ (see
Table~\ref{table:wallexps}).

The wall decorations have the effect of reducing the effective surface
tension of the domains.  As a result, the average domain size  will
decrease.  We can quantify this result by rewriting the free energy of the
 entire strip as
\begin{equation}
\label{eqn:fwt}
{\cal F} = \left(\frac{N}{\Xi_L}\right) \left(2 J L^{d-1} -c_0 h \sqrt{\Xi_L
L^{d-1}} + \delta{\cal F}^{\rm Tot}\right)
.\end{equation}

Here, we used the flat wall result~(\ref{eqn:flatfe}) as a model, but noted
the fact that there is, on average, a further contribution of $\delta{\cal
F}^{\rm Tot}$ per domain
wall from the decorations.

Extremalizing this free energy with respect to $\Xi_L$ gives the average
domain size as
\begin{equation}
\label{eqn:xidec}
\Xi_L = \left(\frac{c_0 h}{J}\right)^{-2}L^{d-1} \left(1+\frac{\delta{\cal
F}^{\rm Tot}}{J L^{d-1}} \right)^2
\end{equation}
As expected, one sees that the domain size is reduced by the effect of
interface roughening.

This correction to the correlation length $\Xi_L$ also has implications for the
phenomenological RG equation.  This equation obtained from~(\ref{eqn:pfss2})
with the use of~(\ref{eqn:xidec}) and~(\ref{eqn:domcorrlat}) is now quite
complex, but it can be simplified by linearizing in $\epsilon$ and using small
$h/J$.  After some tedious algebra, one finds the following RG equation:
\begin{equation}
\label{eqn:eps-hrg}
h^\prime = h \left( 1 + c  h^\gamma \frac{ \ln b}{\ln n} - \frac{\epsilon}{2}
\ln b\right)
\end{equation}
where $c\equiv \tilde{c} c_0^\gamma$ and we have let $J=1$ for simplicity.
Besides the trivial fixed point at $h=0$,~(\ref{eqn:eps-hrg}) has an unstable
fixed point at
\begin{equation}
\label{eqn:eps-hfp}
  h^*  =  \left(\frac{\epsilon \ln n}{2c}\right)^{1/\gamma} +
O(\epsilon^2)^{1/\gamma}
\end{equation}
which remains in the small $h$ region of validity of our description since
$\epsilon$ is regarded as small.
The eigenvalue of~(\ref{eqn:eps-hrg}) is given by
$$
\left.\frac{ \partial h^\prime}{\partial h}\right|_{h=h^*} = 1+\frac{1}{2}
\gamma \epsilon \ln b + {\cal O}(\epsilon^2) = b^{1/\nu}
.$$
This yields the critical exponent
\begin{equation}
  \nu  =  \frac{2}{\epsilon\gamma}
\end{equation}
which characterises the scaling of the bulk correlation length, $\xi \sim | h-
h^\star|^{-\nu}$.
Notice that the arbitrary scale variable $n$ disappears from the critical
exponent $\nu$, which is a universal quantity, though it remains through a weak
logarithmic dependence in $h^\star$ and other non--universal quantities.

If we attempt to let $\epsilon \rightarrow 0$ to extrapolate back to the $h=0$
fixed point, we see that $\nu$ diverges.  This is an indication that the
divergence of the correlation length of the infinite two dimensional system is
faster than any power law.  This divergence can be obtained from
equation~(\ref{eqn:eps-hrg}) at $\epsilon=0$:
\begin{equation}
h^\prime = h \left( 1 + c  h^\gamma \frac{ \ln b}{\ln n} \right)
.\end{equation}
This is equivalent (at small $h$) to
\begin{equation}
\label{eqn:htrans}
\exp\left(\frac{\ln n}{\gamma c (h^\prime)^\gamma}\right) = \frac{1}{b}
\exp\left( \frac{\ln n}{\gamma c h^\gamma}\right)
.\end{equation}
This is the same form as the standard rescaling of a correlation length
$\xi(h^\prime) = b^{-1} \xi(h)$, so the correlation length is
\begin{equation}
\label{eqn:xibulk}
\xi \sim \exp\left( \frac{A}{ h^\gamma}\right)
\end{equation}
close to the fixed point $h^*=0$, where A is the non--universal constant $\ln
n/(c \gamma)$.

The results of these calculations are summarized in
Table~\ref{table:corr-results}.

Here the exponent is $\gamma = 4/3$ for the continuum or $\gamma = 2$ for the
lattice.  The exponential growth of $\xi$ was also found by
Binder~\cite{bin83}.  His calculation was done explicitly on a $2d$ lattice and
he found a form for $\xi$ identical to ours with $\gamma = 2$.

\subsubsection{Conclusions}
\label{sec:lot-conc}

This concludes the $T=0$ investigation of the effects of fluctuations in the
domain wall shape.  The analysis of the RFIM on a lattice and in a continuum
has led to a difference in the critical behaviour of these two cases which
arises from the difference of their
wandering exponents.  This has been seen previously  in other works referred to
above.
One may speculate that this difference is due to the different effect of
surface tension in the lattice versus the continuum.  Interfacial roughening
effectively acts to reduce the surface tension in the domains and is the
crucial factor in breaking the marginality of $h$ in $d=2$.  This surface
tension is coupled to the $d-1$ dimensional area of the domain wall and this is
clearly different between lattice and continuum cases because of the stepped
character necessary on a lattice.  Exactly how this mechanism comes into play
in the RFIM is not clear, however.

The procedure was to perform decorations
on all length scales to find the effect of domain wall roughening on the free
energy and domain size. These determine the phenomenological renormalization
group equation and hence the behaviour near the lower critical dimension.  This
revealed the evolution of the fixed
point as it moves from $h=0$ to $h=h_c \neq 0$ for $d > 2$.

As with the flat wall analysis, both the resulting scaling and the ingredients
which lead to it (size and field dependences of free energy and domain size,
and now also roughening) can be directly checked by numerical work, and this
will be done in section~\ref{sec:numrough}.

In the next section, we
extend our results to include low temperature thermal effects.

\subsection{Low Temperature Theory}
\label{sec:lot-rough}

The effect of the zero temperature wall roughening on the surface tension
is~(by~(\ref{eqn:fwt})) equivalent to the replacement of $J$ by
\begin{equation}
\tilde{J} = J + \frac{\delta{\cal F}^{\rm Tot}}{2 L^{d-1}}
.\end{equation}
Consequently, the low temperature analyses given in sections~\ref{sec:lot}
and~\ref{sec:TMLoT} have to be modified by the replacement $J \rightarrow
\tilde{J}$, in equations~(\ref{eqn:lotfe})--(\ref{eqn:xover})
and~(\ref{eqn:extreme})--(\ref{eqn:TMfeimp}).

The roughening also affects the entropy since it increases the number of
possible configurations.  The entropy change can be obtained crudely by a
slight generalisation of the counting used to derive~(\ref{eqn:binder-d}), or
by a more complete analysis given elsewhere~\cite{omega}.

As remarked in section~\ref{sec:lot}, the zeroth order thermal
scaling~(\ref{eqn:pure-trg}) is not marginal in $d > 1$, and so the
modifications just discussed do not change the RG thermal flow directions for
$h$ small and $d$ near 2.  So the $\epsilon$--expansion results of
section~\ref{sec:zerot-eps} can be combined with those of section~\ref{sec:lot}
to infer the flow diagrams shown in Figure~\ref{fig:rgflow} for $d=1,2,
2+\epsilon$.  For $d=2+\epsilon$ this implies a second order phase transition
at the phase boundary shown, inside which ({\it i.e.} at small $h,T$) long
range order occurs for $h \neq 0$.  This is consistent with previous
renormalization group discussions~\cite{nat88},~\cite{bra85}.

The extended theory outlined in the first two paragraphs above provides the low
temperature free energy modifications produced by rough walls.

This suggests the possibility of distinguishing wall roughening effects by
comparing numerical data on free energy with this theory extended as just
outlined.  This comparison has been made with the transfer matrix free energy
data discussed in section~\ref{sec:TMLoT}.  Despite the excellent quality  of
the data, and its remarkably good agreement with the theory in zeroth order
(flat walls), it was not found possible to extract wall roughening effects from
it.  This is because  the statistical fluctuations in the data, although
extremely small, are nevertheless greater than the roughening contributions in
the ranges of parameters appropriate for the validity of the theory (ensuring
$\xi \gg L$). So, we turn to other more sensitive comparisons.

\subsection{Numerics - Correlation and roughening characteristics from the
Max-flow ground state algorithm}
\label{sec:numrough}

This section describes the extraction of domain wall roughening characteristics
from data obtained using the max--flow algorithm for constructing ground
states.

One comparison uses domain size correlation length data, which is much more
sensitive than the free energy to domain wall roughening, particularly in wide
strips because of the crucial role of roughening in the marginal dimension
$d=2$.  The other involves direct measures of the wall wandering.
These measures are the average over many walls, (for a given $h$ and base scale
$L$) of the maximum height (wandering excursion) $b_{max}$, or of the root mean
square height $b_{rms}$.

Both binary and Gaussian random field distributions were used.  As expected,
the Gaussian distribution gives smoother dependences, and has provided the most
useful data.  But, the binary results are  interesting.
Figure~\ref{fig:GSxibin} gives numerical data  for the configurationally
averaged domain size.  A ``stepping'' of the data as a function of $h$ is
apparent, particularly for smaller $L$.  This is predicted by the flat wall
description~(section~\ref{sec:rw})
equations~(\ref{eqn:xilambda}),~(\ref{eqn:lambda}) since in~(\ref{eqn:lambda}),
$q=\left(2+\lfloor \left(2JL\right)/h \rfloor \right)^{-1}$ involves the
integer part function $\lfloor \; \rfloor$.  There is also a certain
discreteness implied in the use of the transfer matrix.  For comparison, the
flat wall theory is shown in the Figure.  The flat wall random walk analysis
also compares very satisfactorally with corresponding domain size data
(Figure~\ref{fig:gscorr})  from the Gaussian distribution.  Here, the data
contain no steps.  If one repeats the random walk analysis with a Gaussian
random field distribution, the ``integer part'' function does not occur, and
the transfer matrix becomes a convolution operator.  Since the eigenfunctions
are again plane waves, the final result for the correlation length is the same,
except that it has no ``integer part'' function, and thus no ``stepping''.
But, though the Gaussian distribution must therefore set the large length scale
behaviour, in energy (Figure~\ref{fig:fefit1}), domain size (above) and
probably in time scales for relaxing on given characteristic scales
(``dynamics''), there can be real differences for these quantities between
binary and Gaussian cases at smaller scales.

The domain size data can be fitted to the flat wall theory or to the analysis
including wall roughening (sections~\ref{sec:decondec},~\ref{sec:zerot-eps}).
Equation~(\ref{eqn:xiflat}) gives the basic flat wall description, in which for
strips ($d=2$), $\Xi_L \sim L h^{-2}$, and one fitting coefficient ($c_0$)
occurs.  The fuller random walk version of the flat wall theory
(section~\ref{sec:rw}) is equivalent, for small $h/L$, but more generally it
gives a $\Xi_L$ different by a ``discreteness factor'' $(1+{\cal O}(h/L))$.
The wall roughening analysis instead corrects the flat wall $\Xi_L$ by a factor
$(1+{\cal O}(h^\gamma L^{\theta - d+1} \ln L))$.  This should be amalgamated
with the discreteness  factor evident from the RW analysis.  Then the domain
wall roughening theory gives the best quality of fit to the max--flow data on
$\Xi_L$.  For instance, considering a Gaussian random field distribution,
Table~\ref{table:xifit} shows the decreased statistical error $\chi^2$
associated with the addition of this domain roughening term.  Even though this
extra term provides an extra free constant to the fit ($c=\tilde{c}
c_0^\gamma$), it does not necessarily decrease $\chi^2$.  Indeed, it worsens
the fit to the zero'th order theory since the higher order ``discreteness
term'' has been neglected.  When this term is ignored, the roughening only has
a correlation coefficient of $r=0.218$.  However, when this discreteness term
has been included from the more complete random walk analysis, the wall
roughening is seen to decrease the value of $\chi^2$.  It then has a much more
significant correlation coefficient of $r=0.816$.

The max--flow data in the wall decoration variables $b_{rms}$, $b_{max}$ give
much more conclusive evidence for the roughening effects and provide
quantitative estimates of exponents.    Data for the $h$--dependence of
$b_{rms}$ for $L=8$ are shown in Figure~\ref{fig:gscomp} for both binary and
Gaussian distributions.  Again, the jagged character of the binary result is
evident, even at this moderately large $L$.  So the remaining discussion is
confined to the Gaussian case.  To compare with the
prediction~(\ref{eqn:b-general}), a log--log plot of the Gaussian $b_{rms}$
versus $h$ is given for $L=2,\ldots,11$ in Figure~\ref{fig:gsloglog}.  A very
similar plot is obtained also for $b_{max}$, though the absolute size of
$b_{max}$ is larger, by up to a factor of $4$.  The departure from linearity of
the log--log plot at small $h$ is almost certainly a lattice effect.  That
requires $b \lower3pt\hbox{$\, \buildrel > \over \sim \, $} 1$, and using {\it
e.g.}~(\ref{eqn:blat}) (with $c_0 = 1.8$) to estimate where $b \approx 1$
suggests that~(\ref{eqn:b-general}) should break down for $h/J
\lower3pt\hbox{$\, \buildrel < \over \sim \, $} 5 L^{-1/2}$, which is in
qualitative agreement with what is seen.
Similarly, a breakdown of the simple theory giving~(\ref{eqn:b-general}) is
expected when $b$ (strictly $b_{max}$) becomes comparable to $\Xi_L$, {\it
i.e.} (using~(\ref{eqn:blat}) and~(\ref{eqn:xiflat})) for $h/J
\lower3pt\hbox{$\, \buildrel > \over \sim \, $} 1.5$.  This $L$--independent
cutoff is consistent with the departures from linearity seen at the largest
fields in the log--log plot.  The scaling window between the upper and lower
$h$ departures is quite wide for the largest $L$'s, and consequently they
should give the most accurate field exponent $\kappa$ from power law fits
to~(\ref{eqn:b-general}) within the scaling window.  The results, from both
$b_{rms}$ and $b_{max}$ are shown in Figure~\ref{fig:gsexp} and are consistent
with $\kappa \approx 2.1 \pm 0.3$.  Figure~\ref{fig:gscoeff} shows the
$L$--dependent prefactor of $h^\kappa$ in $b_{rms}$ and $b_{max}$.  In each
case, the prefactor is very close to linear in $L$ in the larger $L$ scaling
regime, consistent with $\zeta=1$ in~(\ref{eqn:b-general}).  The behaviours
seen in Figure~\ref{fig:gsloglog}, and in the corresponding plot for $b_{max}$,
as well as the associated exponents $\kappa, \zeta$ are therefore consistent
with the analysis in section~\ref{sec:rough-theory}.

\section{Conclusions}
\label{sec:final-conc}

In this paper we have developed a domain scaling description for the random
field Ising model by exploiting a bar geometry.  As in earlier work of
Villain~\cite{vil82}, Grinstein and Ma~\cite{gri82} and Binder~\cite{bin83},
the marginality of the basic flat wall picture at the lower critical dimension
$d_l = 2$ is removed by wall roughening effects~(Section~\ref{sec:rough}).  The
resulting phenomenological renormalization group transformation has been used
to obtain the critical properties.  In particular, the special critical
behaviour  of the two--dimensional correlation length and the phase diagram in
$d=2+\epsilon$ have been given~(Section~\ref{sec:zerot-eps}).

The wall roughening ingredient has its own scaling characteristics (the
exponents $\gamma,\theta,\kappa,\zeta$) which have been here discussed using
both a simple analytic method (following Natterman~\cite{nat85}) and a field
theoretic approach.  These exponents have a direct bearing upon the RG
transformation and hence on the scaling properties.

An essential element to this study has been the support that numerical studies
have provided for the analytic description.  The comparison between numerical
data and analytic predictions was made in the strip geometry so that the
critical ingredients of the theory could be directly investigated.  These
ingredients include the domain size, the free energy, and the roughening
characteristics, as functions of $h,T$ and strip width $L$.

The numerical approaches   employed were transfer matrix and Monte Carlo
techniques, and a ground state (max--flow) algorithm.  The first two give free
energy and domain size data confirming the basic flat wall picture through:
$(i)$ the $h^2/L$ dependence of the $T=0$ flat wall energy at low $h$, and its
low temperature generalisation; $(ii)$ the domain size distribution and the
correlation length--domain size relationship.

The numerical ground state data, obtained via the max--flow algorithm, directly
verify the basic validity of the domain size predicted by the (flat wall)
random walk analysis, but they also provide evidence for the roughening effects
in the domain size.  The max--flow data on the wall decoration variables
$b_{rms}$, $b_{max}$ are the most conclusive evidence for roughening effects.
They give power law scaling in a window of the size predicted by the theory,
sufficient at large $L$ to give quite accurate values for roughening exponents,
in good agreement with the (lattice based) theory.

The conclusion is that the theory contains the correct ingredients, and the
numerical data provide quantitative confirmations of the way they enter into
the theory.

We have not made corresponding numerical investigations of the $d=3$ case,
where it is of course more difficult to obtain data for the range of $L$'s we
needed here.  However, we hope the present work may stimulate such an effort,
which ({\it e.g.} using the ground state algorithm) could sort out the so far
unresolved critical behaviour~\cite{you93,gof93}.  Another extension, presently
under consideration, is to the kinetic behaviour, for which the free energy
scaling discussed here is an essential ingredient.



\begin{figure}
\caption[]{The bar geometry, showing the flat domain
walls that separate regions of up and down spins. The
average domain size $\Xi_L$ is the average of
the $\Xi_i$'s.}
  \label{fig:stripgeom}
\end{figure}

\begin{figure}
\caption[]{Illustration of how the random field configuration may be viewed
as a random walk in $1+1$ dimensions.  We use $L=3$ here. }
  \label{fig:rw}
\end{figure}

\begin{figure}
  \caption{Log--log plots of the excess free energy at $T = 0.1 J$.  }
  \label{fig:TMfefit}
\end{figure}

\begin{figure}
\caption[]{A comparison of the free energy as determined by the transfer
matrix data, and that determined by our low temperature theory using the
fitted constant $c_0 = 3.406$, at $T=0.1 J$ and $T=0.3J$.}
\label{fig:fefit1}
\end{figure}

\begin{figure}
\caption[]{The free energy as a function of $h$ for $T/J = 0.1$ and
$L\in(2,7)$.  These data were determined by the transfer matrix method, and
illustrate the three regimes of $h$ dependence}
\label{fig:TMhrange}
\end{figure}

\begin{figure}[hbtp]
  \caption[]{The average energy of the $L=1$ and $L=4$ RFIM, as determined by
both the Monte Carlo algorithm (data points) and the zeroth order theory
(lines).  The symbol $\Diamond$ corresponds to $h=0$, $-$ to $h=0.1J$, $\Box$
to $h=0.2J$ and $\times$ to $h=0.5J$.}
\label{fig:mcavel14}
\end{figure}

\begin{figure}[hbtp]
 \caption[]
{The correlation length $\xi_L$ plotted against the domain size $\Xi_L$ for
$L=2$.  We also include the predicted relation between these quantities for
comparison.  Plots for $L=1,4$ are essentially the same.}
\label{fig:mccorrl24}
\end{figure}

\begin{figure}[hbtp]
   \caption[]{The distribution of domain sizes according the the measure
$\Xi_L$ as found by the Monte Carlo routine when $L=1$.}
  \label{fig:mcdomdistl1}
\end{figure}

\begin{figure}[hbtp]
  \caption[]
{The domain size $\Xi_L$ plotted as a function of temperature for $L=1$ and
$L=4$.  The data points represent the MC data whereas the lines show the result
of solving the flat wall theory.  The symbol $\Diamond$ corresponds to $h=0$,
$-$ to $h=0.05J$, $\Box$ to $h=0.1J$, $\times$ to $h=0.2J$, and $\triangle$ to
$h=0.5J$.  The flattening of the data at low $T$ is due to the finite length of
the simulated system.}
  \label{fig:mcdoml24}
\end{figure}

\begin{figure}\caption{The typical domain size at $T=0$ from the max--flow
algorithm.  The points represent the data from the max--flow algorithm, and the
solid lines represent the theoretical form with a fitted multiplicative
constant.}
\label{fig:gscorr}
\end{figure}

\begin{figure}   \caption[]{First order decorations to a domain wall on a
lattice (left) and
in a continuum (right).  Here, $d=3$.
        The decorations have height $b$ and width $a$.}
   \label{fig:balat}
\end{figure}

\begin{figure}  \caption{Plot of the free energy exponent $\theta$ and the
wandering exponent
  $\zeta$ for the wall decorations on a lattice and in a continuum.}
  \label{fig:expplot}
\end{figure}

\begin{figure}
   \caption[]{The process of placing decoration upon decorations, such that
decorations occur on all length scales.  This process has a cut--off length
scale when the size of the decorations become on order of the lattice
constant.}
  \label{fig:decondec}
\end{figure}

\begin{figure}[h]
  \caption[]{The RG flow diagram for the RFIM as predicted by our finite size
scaling analysis. }
  \label{fig:rgflow}
\end{figure}

\begin{figure}[h]
  \caption[]{The domain size as a function of $h/J$ at $T=0$ determined by the
max--flow algorithm (data points), and the flat wall analysis (lines) for a
binary field distribution.}
  \label{fig:GSxibin}
\end{figure}

\begin{figure}  \caption{The wall roughening exponent $\kappa$ as determined
from power law fits to $b_{rms}$ and $b_{max}$.  For large $L$, these tend
towards $2$.}
  \label{fig:gsexp}
\end{figure}

\begin{figure}  \caption{The wall roughening term $a^\zeta$ as determined from
power law fits to $b_{rms}$ and $b_{max}$.  For large $L$, these tend towards a
linear dependence on $L$, consistent with $\zeta=1$, as shown by the lines of
best fit.}
  \label{fig:gscoeff}
\end{figure}

\begin{figure}  \caption[]{The root--mean--square domain wall width as a
function of $h/J$ for various length scales $L$.  The use of a log--log plot
shows the scaling window where the plots are linear.}
  \label{fig:gsloglog}
\end{figure}

\begin{figure}  \caption[]{The rms domain wall width as a function of $h/J$
for $L=8$.  This plot emphasizes the difference between the binary and Gaussian
random field distributions, showing the stepped nature of the former as more
excitations become favourable at lower $h/J$.}
  \label{fig:gscomp}
\end{figure}


\begin{table}[hbtp]
\caption{Fits to the free energy data from the transfer matrix}
\label{table:Q}
\begin{center}
\begin{tabular}{lddd}
L & Slope & Error& $Q$  \\ \hline
2 & 1.792 &  0.007 & 0.0004 \\
3 & 1.848 &  0.009 & 0.33 \\
4 & 1.857 &  0.009 & 0.79 \\
5 & 1.885 &  0.011 & 0.96 \\
6 & 1.901 &  0.012 & 0.95 \\
7 & 1.904 &  0.012 & 1.00 \\
8 & 1.945 &  0.015 & 1.00 \\
9 & 1.958 &  0.014 & 1.00 \\
\end{tabular}
\end{center}
\end{table}

\begin{table}
\caption{Fits of the free energy to the transfer matrix data at finite
temperature}
\label{table:Q2}
\begin{center}
\begin{tabular}{lldll}
 L  & {\cal N} &  $c_0$  & $\chi^2$ & $Q$ \\ \hline
2 - 9 & 120    &  1.774 &  457     &  $10^{-41}$ \\
2 - 7 & 90     &  1.761 &  375     &  $10^{-37}$ \\
2 - 5 & 60     &  1.745 & 288      &  $10^{-32}$ \\
4 - 9 & 90     &  1.846 & 81    & 0.686  \\
\end{tabular}
\end{center}
\end{table}

\begin{table}
\caption{Main results of the Flat Wall analysis}
\label{table:corr-results}
\begin{center}
\begin{tabular}{ccc}
  Dimensions   & Fixed Point     & Bulk Correlation Length \\ \hline
2              &     $h^* = 0$   & $\xi \sim \exp \left(A/ h^\gamma\right)$ \\
$2 + \epsilon$ & $h^* = \left(\frac{\epsilon \ln n}{2c}\right)^{1/\gamma}$ &
$\xi \sim \left( h-h^*\right)^{ \frac{-2}{\epsilon \gamma} }$ \\
\end{tabular}
\end{center}
\end{table}

\newpage
\mediumtext

\begin{table}
\caption{Comparison of Approaches to Wall Decoration Problem}
\label{table:wallexps}
\begin{center}
\begin{tabular}{lllllll}
 & \multicolumn{3}{c}{$d < 2$} &
\multicolumn{3}{c}{$d \geq 2$} \\
\cline{2-4} \cline{5-7}
        & Lattice & Simple     & Field  &Lattice & Simple        & Field \\
        &         & Continuum  & Theory &        & Continuum     & Theory \\
\hline
$\zeta$ & $3-d$  & $3-d$       & N.A.   &$3-d$   &$(5-d)/3$     &$(5-d)/3$\\
$\theta$&$3-d$  & 1       & N.A.   &$3-d$   &$(d+1)/3$      &$(d+1)/3$\\
$\gamma$& 2     & 2       & N.A.   &2       &$4/3$          &$4/3$    \\
$\kappa$& 2     & 2       & N.A.   &2       &$2/3$          &$2/3$    \\
\end{tabular}
\end{center}
\end{table}

\begin{table}
\caption{Results of fits to the $T=0$ domain size data}
\label{table:xifit}
\begin{center}
\begin{tabular}{lllll}
 & \multicolumn{2}{c}{Flat Domain Walls} &
\multicolumn{2}{c}{Roughened Domain Walls} \\
\cline{2-3} \cline{4-5}
 & Zeroth Order & Random Walk & Zeroth Order & Random Walk\\
$\chi^2$ & 4.215 & 0.607 & 4.447 & 0.404
\end{tabular}
\end{center}
\end{table}


\begin{thebibliography}{10}

\bibitem{imr75}
Y. Imry and S. Ma, Phys. Rev. Lett. {\bf 35},  1399  (1975).

\bibitem{gri76}
G. Grinstein, Phys. Rev. Lett. {\bf 37},  944  (1976).

\bibitem{aha76}
A. Aharony, Y. Imry, and S. Ma, Phys. Rev. Lett. {\bf 37},  1364  (1976).

\bibitem{you77}
A. Young, J. Phys. A {\bf 10},  L257  (1977).

\bibitem{aha78}
A. Aharony, Phys. Rev. B {\bf 18},  3318  (1978).

\bibitem{aha78b}
A. Aharony, Phys. Rev. B {\bf 18},    (1978).

\bibitem{vil82}
J. Villain, J. Phys. (Paris) {\bf 43},  L551  (1982).

\bibitem{gri82}
G. Grinstein and S. Ma, Phys. Rev. Lett. {\bf 49},  685  (1982).

\bibitem{imb84}
J. Imbrie, Phys. Rev. Lett. {\bf 53},  1747  (1984).

\bibitem{bri87}
J. Bricmont and A. Kupiainen, Phys. Rev. Lett. {\bf 59},  1829  (1987).

\bibitem{aiz89}
M. Aizenman and J. Wehr, Phys. Rev. Lett. {\bf 62},  2503  (1989).

\bibitem{nat88}
T. Nattermann and J. Villain, Phase Transitions {\bf 11},  5  (1988).

\bibitem{you93}
H. Rieger and A. Young, J. Phys. A {\bf 26},  5279  (1993).

\bibitem{gof93}
J. Gofman {\it et~al.}, Phys. Rev. Lett. {\bf 71},  1569  (1993).

\bibitem{fis79}
S. Fishman and A. Aharony, J. Phys. C {\bf 12},  L729  (1979).

\bibitem{yos82}
H. Yoshizawa {\it et~al.}, Phys. Rev. Lett. {\bf 48},  438  (1982).

\bibitem{bel83}
D. Belanger, A. King, V. Jaccarino, and J. Cardy, Phys. Rev. B {\bf 28},  2522
  (1983).

\bibitem{sha84}
Y. Shapira, N.~O. Jr., and S. Foner, Phys. Rev. B {\bf 30},  6639  (1984).

\bibitem{bir85}
R. Birgeneau, R. Cowley, G. Shirane, and H. Yoshizawa, Phys. Rev. Lett. {\bf
  54},  2147  (1985).

\bibitem{yos85}
H. Yoshizawa, R. Cowley, G. Shirane, and R. Birgeneau, Phys. Rev. B {\bf 31},
  4538  (1985).

\bibitem{mit86}
P. Mitchel {\it et~al.}, Phys. Rev. B {\bf 34},  4719  (1986).

\bibitem{ges88}
Geschwind and Ogielski, J. Appl. Phys. {\bf 63},  3291  (1988).

\bibitem{ram88}
C. Ramos, A. King, and V. Jaccarino, Phys. Rev. B {\bf 37},  5483  (1988).

\bibitem{fer91}
I. Ferreira, A. King, and V. Jaccarino, Phys. Rev. B {\bf 43},  10797  (1991).

\bibitem{led92}
M. Lederman {\it et~al.}, Phys. Rev. Lett. {\bf 68},  2086  (1992).

\bibitem{bec93}
C.~C. Becerra and A. Paduan-Filho, J. Appl. Phys. {\bf 73},  5491  (1993).

\bibitem{hil93}
J. Hill, Q. Feng, R. Birgeneau, and T. Thurston, Phys. Rev. Lett. {\bf 70},
  3655  (1993).

\bibitem{hil91}
J. Hill {\it et~al.}, Phys. Rev. Lett. {\bf 66},  3281  (1991).

\bibitem{gri83b}
G. Grinstein and S. Ma, Phys. Rev. B {\bf 28},  2588  (1983).

\bibitem{bin83}
K. Binder, Z. Phys. B. {\bf 50},  343  (1983).

\bibitem{vil85}
J. Villain, J. Physique {\bf 46},  1843  (1985).

\bibitem{nat85}
T. Natterman, Phys. Stat. Sol. (b) {\bf 131},  563  (1985).

\bibitem{intro}
R. Stinchcombe, E. Moore, and S. de~Queiroz, preprint (unpublished).

\bibitem{mymc}
E. Moore and R. Stinchcombe, preprint (unpublished).

\bibitem{ogi86a}
A. Ogielski, Phys. Rev. Lett. {\bf 57},  1251  (1986).

\bibitem{far93}
E. Farhi and S. Gutmann, preprint submitted to Phys. Rev. B (unpublished).

\bibitem{pri89}
V. Privman and N. \v{S}vraki\'c, {\em Directed Models of Polymers, Interfaces
  and Finite--Size Properties}, Vol.~338 of {\em Lecture Notes in Physics}
  (Springer, Berlin, 1989).

\bibitem{que92}
S. de~Queiroz and R. Stinchcombe, Phys. Rev. B {\bf 46},  6635  (1992).

\bibitem{tm17}
A. Crisanti, S. Nicholis, G. Paladin, and A. Vulpiani, J. Phys. A {\bf 23},
  3083  (1990).

\bibitem{tm18}
B. Derrida, Phys. Rep. {\bf 103},  29  (1984).

\bibitem{mitreport}
A. Goldberg, Internal report MIT/LCS/TM-291, Laboratory for Computer Science,
  Massachusetts Institute of Technology (unpublished).

\bibitem{mapping}
J. Picard and H. Radcliffe, Networks {\bf 5},  357  (1974).

\bibitem{numrec}
W. Press, B. Flannery, S. Teukolsky, and W. Vetterling, {\em Numerical Recipes
  in C, The Art of Scientific Computing}, second edition ed. (Cambridge
  University Press, ADDRESS, 1994).

\bibitem{gri84}
G. Grinstein, J. Appl. Phys. {\bf 55},  2371  (1984).

\bibitem{par79}
G. Parisi and N. Sourlas, Phys. Rev. Lett. {\bf 43},  744  (1979).

\bibitem{mez90}
M. M\'ezard and G. Parisi, J. Phys. A {\bf 23},  L1229  (1990).

\bibitem{mez92}
M. M\'ezard and A. Young, Europhys. Lett. {\bf 18},  653  (1992).

\bibitem{fer83}
J. Fernandez, G. Grinstein, Y. Imry, and S. Kirkpatrick, Phys. Rev. Lett. {\bf
  51},    (1983).

\bibitem{omega}
E. Moore, preprint (unpublished).

\bibitem{bra85}
A. Bray and M. Moore, J. Phys. C {\bf 18},  L927  (1985).

\end{thebibliography}
\end{document}